\documentclass[preprint,showpacs,preprintnumbers,amsmath,amssymb]{revtex4}
\usepackage{graphicx}
\usepackage{dcolumn}
\usepackage{bm}
\begin{document}

\title{Characterization of complex networks by higher order neighborhood properties }
\author{Roberto F. S. Andrade, Jos\'e G. V. Miranda, Suani T. R. Pinho}
\affiliation {Instituto de F\'{\i}sica - Universidade Federal da Bahia \\
40.130-240 - Salvador - Brazil}
\author{ and Thierry Petit Lob\~{a}o}
\affiliation{
Instituto de Matem\'{a}tica -  Universidade Federal da Bahia,\\
40210-340 - Salvador - Brazil}

\date{\today}

\begin{abstract}

A concept of higher order neighborhood in complex networks,
introduced previously (PRE \textbf{73}, 046101, (2006)), is
systematically explored to investigate larger scale structures in
complex networks. The basic idea is to consider each higher order
neighborhood as a network in itself, represented by a corresponding
adjacency matrix. Usual network indices are then used to evaluate
the properties of each neighborhood. Results for a large number of
typical networks are presented and discussed. Further, the
information from all neighborhoods is condensed in a single
neighborhood matrix, which can be explored for visualizing the
neighborhood structure. On the basis of such representation, a
distance is introduced to compare, in a quantitative way, how far
apart networks are in the space of neighborhood matrices. The
distance depends both on the network topology and the adopted node
numbering. Given a pair of networks, a Monte Carlo algorithm is
developed to find the best numbering for one of them, holding fixed
the numbering of the second network, obtaining a projection of the
first one onto the pattern of the other. The minimal value found for
the distance reflects differences in the neighborhood structures of
the two networks that arise from distinct topologies. Examples are
worked out allowing for a quantitative comparison for distances
among a set of distinct networks.
\end{abstract}


\maketitle

\section{INTRODUCTION}

In recent years, complex networks have become one of the major
research themes in complex systems. They allow for an interesting
interplay between mathematics and physics through the
well-established theories of graphs \cite{graphtheory_book} and
statistical mechanics. Despite the fact that they are simply defined
in terms of sets of $N$ nodes and $L$ links connecting pairs of
nodes, they can exhibit a very complex structure. In this way, they
have attracted the attention of scientists, from most different
areas, aiming to investigate their potential usefulness in the
description of geometrical and dynamical properties of complex
systems \cite{Dorogovtsev,Boccaletti,networks_book}.

The huge development in this research area comes along with the
proposition of parameters that capture some essential properties
that such objects may have. Most of the current works in the
literature expresses the quantitative results in terms of a hand
full of indices or quantifiers \cite{Barabasi_review, Newman_review,
Newman_assortativity}, like the average number of links per node
$\langle k\rangle$, clustering coefficient $C$, mean minimal
distance among the nodes $\langle d\rangle$, diameter $D$, and the
assortativity degree $A$. Also of relevance are certain relations
that express how these quantifiers are composed, like $p(k)$, the
probability distribution of nodes with $k$ links, the distribution
of each individual node clustering coefficients $C(k)$ with respect
to its node degree, and the average degree of the neighbors of a
node with $k$ links $\langle k_{nn}(k)\rangle$.

The quoted set of indices have been used as they can be easily evaluated,
and have a very clear intuitive geometrical and topological meaning.
Despite their ability in characterizing these most relevant features of
networks, other parameters can be proposed to account for other properties
that are overlooked by the above quoted set. The actual relevance of the
information brought by a new parameter depends whether it is, in a certain
sense, \textit{orthogonal} to the set of parameters that had been used so
far. In such case, its evaluation contributes either to emphasize
similarities among a set of networks, or to uncover differences among
them. Therefore, new parameters may indeed become relevant to identify
differences between networks that could have been thought to be similar in
a more restricted space of characterization parameters.

From our point of view, it is important to probe the network with
respect not only to the immediate, close neighborhood of a node, but
also how all pairs of nodes are related among themselves, from the
nearest up to the maximal distance $D$. This leads to the concept of
higher order neighborhoods, providing a clearer picture on the
intermediate and large scale structures of the network. It is worth
mentioning that other authors \cite{Fronczak, Costa, Brunet,
Caldarelli} also defend similar positions to ours, having proposed
several strategies to investigate network properties that advance
beyond the immediate vicinity of a node. However, as it shall be
clear in the sequel, our approach is quite different from those just
presented in the related literature.

In a previous work \cite{Andrade_neighborhood}, we have indicated
how to obtain all higher order neighborhoods of a node in a
network $R$ with the help of Boolean product of matrices.
According to the definition introduced there, two nodes in $R$ are
neighbors of order $O(\ell), \ell=1, 2, ..., D$, when the shortest
path connecting them, along links in $R$, has $\ell$ steps. Each
$O(\ell)$ neighborhood of $R$ determines a network $R(\ell)$, just
by connecting pairs of nodes that are $\ell$ steps apart in $R$.
Each $R(\ell)$ is fully described by the corresponding adjacency
matrix $M(\ell)$. This process has three main consequences: i) it
unfolds, in a straightforward way, a large amount of information
contained in the adjacency matrix $M$ of $R$; ii) doing this, it
puts the information at hand for the analysis of the higher order
neighborhood properties; iii) it offers, quite naturally, to
condense all information extracted from $M$ in a single
neighborhood matrix  $\mathbf{\widehat{M}}$, defined in terms of
the set $\{M(\ell)\}$, which can be used for a much complete
characterization of $R$ than $M$. To attest the consistency and
usefulness of the proposed framework, that work also presented a
characterization of the $O(\ell)$ neighborhoods by means of the
eigenvalue spectra of the higher order adjacency matrices
$M(\ell)$, for several types of networks.

The current work has two main purposes, both of which are related to
developments of the quoted procedure and, to our opinion, they
contribute to provide a more complete insight into the properties of
the analyzed networks. The first one is to characterize the larger
structure properties by the analysis of all $O(\ell)$ neighborhoods
of $R$. Like in the analysis of the spectral density, the method
amounts to analyze all networks $R(\ell)$, taking into account their
representation through the corresponding $M(\ell)$.  As they are
\textit{bona fide} networks, their immediate neighborhood can be
analyzed with the same set of indices used to characterize $R$,
i.e., $\langle k (\ell)\rangle$, $C(\ell)$, $p(k;\ell)$, $A(\ell)$,
and the distribution $C(k;\ell)$. The exact values of the global
network parameters $D$ and $\langle d\rangle$ are readily obtained
after the decomposition process. In addition to that, the adopted
procedure provides, as by-product, the necessary measures to
evaluate the fractal dimension $d_F$ of a network, according to the
scheme \cite{Havlin} proposed recently.

In regard to the second series of results, we first use the matrices
$\mathbf{\widehat{M}}$ for the purpose of visualizing the large
scale structure of the networks. Afterwards, they are systematically
explored to define and evaluate a measure of the distance between
two networks with the same number of nodes. This implementation
provides a still more direct comparison between networks than that
obtained by the comparison of a larger number of indices as
discussed above. Such procedure would not be so efficient if defined
only on the basis of the original adjacency matrix $M$, as it
requires the information on the complete neighborhood structure
present in $\mathbf{\widehat{M}}$.

The rest of this work is so organized: in Section II we briefly
review the main steps required to describe the higher order
networks, and discuss how the quantities obtained in this process
can be used to describe and represent other properties of networks.
In Section III we discuss the behavior of $C(\ell)$, $\langle
k(\ell)\rangle$, $p(k;\ell)$, $C(k;\ell)$, $A(\ell)$, and $d_F$ for
a large number of networks. We consider the small-world (SW)
\cite{Watts} and scale-free (SF), generated according, respectively,
to the Newman \cite{Newman_review} and the Barabasi-Albert
\cite{Barabasi_science} algorithms, the Erd\"os-Renyi (ER)
\cite{Renyi}, the Cayley tree (CT), the Apollonian network (AN)
\cite{Andrade_apollonian}. We also include in this list two well
known structures, the Diamond Hierarchical Lattice (DHL) and the
Wheatstone Hierarchical Lattice (WHL), which have been largely used
in the context of spin models in statistical mechanics
\cite{Griffiths}. A comparison of the values for $C(\ell)$ with
similar parameters proposed in other works to characterize the
topology of more distant neighbors is included \cite{Newman_review,
Fronczak, Caldarelli}. In Section IV, we first show how a graphical
color representation of the distinct neighborhoods can be worked
out. This issue is followed by a discussion on the effect of node
numbering in the network representation: we propose two renumbering
procedures and a measure to estimate the distance, based on the
neighborhood structure, between two networks with the same number of
nodes. Results to this investigation are presented in Section V, for
a subset of networks used in Section III. Finally, in Section VI, we
close the work with final remarks and conclusions.

\section{EVALUATING $O(\ell)$ NEIGHBORHOODS}

The $O(\ell)$ neighborhoods can be easily evaluated if we
represent a network $R$ by its adjacency matrix $M$. The networks
$R(\ell)$, defined in the Introduction, are expressed by the
corresponding adjacency matrix $M(\ell)$. So, if $O_i(\ell)$ is
the set of $\ell$-neighbors of node $i$, we obtain
\begin{equation}\label{eq1}
M(\ell)_{ij} = \left \{
\begin{array}{ll}
  \delta_{\ell,\ell'}, &  $if $ j \in O_i(\ell')  \\
  0, & $otherwise$
\end{array}
\right.
\end{equation}
We note that $M=M(1)$ and, if we consider that each node belongs
to $O(0)$ of itself, then $M(0)=I$, where $I$ represents the
identity matrix. As shown in \cite{Andrade_neighborhood}, all
${M(\ell)}$ can be successively evaluated with the help of Boolean
operations \cite{Boolean}, using one single recurrence equation:
\begin{equation}\label{eq2}
M(\ell)=(\bigoplus_{g=0}^{\ell-1}M(g))\otimes M(1) -
(\bigoplus_{g=0}^{\ell-1}M(g)).
\end{equation}
The evaluation of all $M(\ell)$'s opens the path for a direct
evaluation of many network parameters. Many of them are neighborhood
dependent and, as far as we know, have not been considered before.
On the other hand, the evaluation of some global parameters, which
have usually been performed along other methods, can be obtained
within this framework in a straight forward way. Let us first note
that, once $\textit{N}$ is finite, for some large enough
$\ell_{max}$ we find $M(\ell)\equiv 0, \forall \ell>\ell_{max}$, so
that $D\equiv\ell_{max}$. Next, the knowledge of all $M(\ell)$
allows for the definition of the matrix $\mathbf{\widehat{M}}$,
which carries all information on the shortest path between any two
vertices $i$ and $j$ along the network. As, according to
(\ref{eq1}), $M(\ell)_{ij}=1$ for only one $\ell$, the definition
\begin{equation}\label{eq3}
\mathbf{\widehat{M}}=\sum_{g=0}^{D}g M(g),
\end{equation}
implies that, for all pairs $(i,j)\in O(\ell)$,
$\mathbf{\widehat{M}}_{ij}=\ell$. Using color or gray code plots,
the network neighborhood structure entailed in
$\mathbf{\widehat{M}}$ can be visualized. The matrix
$\mathbf{\widehat{M}}$ is only a bit differently defined from the so
called \textit{distance matrix} \cite{Harary} used in the graph
theory; it is reduced to that one in the case of a connected
network. With the help of (\ref{eq3}), the average minimal path for
a node $i$ is easily expressed by
$d_i=(\sum_{j=1}^{N}\mathbf{\widehat{M}}_{i,j})/(N-1)$, leading,
immediately, to the average $\langle d\rangle$.

Despite the fact that several complex networks have an intrinsic
length scale, represented e.g., by $\langle d\rangle$ or $D$, there
have been some attempts to associate fractal dimensions to these
objects. The definition for a fractal dimension $d_F$, cited in the
introduction, is deeply related to the concept of $O(\ell)$
neighborhoods. Indeed, once length is measured in $R$ by the number
of steps between nodes, scaling arguments lead to $L(\ell)\sim \ell
^{d_F}$, where $L(\ell)$ counts the number of pairs of nodes that
are $\ell$ steps apart. Within the current approach we obtain, at
once, $L(\ell)=\sum_{i,j=1}^{N}M(\ell)_{i,j}/2$.

The evaluation of large scale structures in the network proceeds by
considering each $R(\ell)$, represented by $M(\ell)$, as an
independent network. Therefore, the parameters $C(\ell)$, $\langle
k(\ell)\rangle$, $p(k;\ell)$, $C(k;\ell)$ and $A(\ell)$, which
describe the local information on each $R(\ell)$, also provide
information on the large scale structures of $R$. For any value of
$\ell$ and node $i$, the node degree
$k_i(\ell)=\sum_{j=1}^{N}M(\ell)_{i,j}$ and the node clustering
coefficient $C_i(\ell)=\sum_{m\in
O_i(\ell)}\sum_{j=1}^{N}M(\ell)_{i,j}M(\ell)_{m,j}/2$ are directly
expressed in terms of elements of $M(\ell)$, while the other three
quantities follow immediately by counting the number of occurrences
of nodes with degree $k(\ell)$.

To characterize a data set represented as a network, one has to know
whether all points are indeed connected among themselves in a single
component, or partitioned into disjoint sub-networks. This important large
scale property can be exactly answered provided the set ${M(\ell)}$ is
evaluated. If the network consists of a single component, the quantity
$Z$, defined by
\begin{equation}\label{eq4}
    Z\equiv\sum_{\ell = 1}^{D}\sum_{j=1}^{N}M(\ell)_{i,j},
\end{equation}
always assumes the value $N(N-1)$. If $Z<N(N-1)$, two or more components
are present and, in this case, their number and corresponding sizes can be
evaluated as follows. First evaluate
$\kappa_i=1+\sum_{\ell=1}^{D}k_i(\ell)<N-1, \forall i$. It counts the
number of nodes in the specific component the node $i$ belongs to, but it
also indicates possible sizes for any other component. The maximal number
of distinct values assumed by $\kappa_i$ is limited by
$(-1+(1+8N)^{1/2})/2$. If $\sigma(\kappa)$ represents the number of nodes
that share the same value of $\kappa$, the number of components of this
size is simply $\sigma(\kappa)/\kappa$, what completes the
characterization on the partition of the network. The adjacency matrix of
a non connected network, by a suitable rearrangement of its nodes, may be
reduced to a form of non-zero diagonal blocks, i.e. a direct sum of
matrices of smaller order, each block corresponding to a connected
component of the network. Because of this, one may just deal with
connected networks, as we shall consider herein. We would like to
emphasize that, for the sake of simplicity, we will restrict ourselves to
present results only for undirected networks, without self links, and
parallel links between any two nodes.

\section{NEIGHBORHOOD CHARACTERIZATION}

In this Section we present a characterization of $O(\ell)$
neighborhood for some standard networks. A flavor of this
procedure is available in a previous work
\cite{Andrade_neighborhood}, where we concentrated on the spectral
properties of the networks $R(\ell)$.

We have obtained a large amount of information on the neighborhood
structure of networks. A summary of our most interesting results are
depicted in Figures 1-6, where we draw the parameters quoted in the
previous Section as function of the neighborhood $\ell$. As anticipated in
the Introduction, for the sake of a clearer discussion of our results, we
use the methodology to investigate both geometrically grown networks (CT,
AN, DHL, WHL), as well as many examples of well known networks, generated
by precisely defined algorithms based on random generators, like ER, SF
and SW networks. We discuss results for networks ranging from $N\sim100$
up to a maximum of $N=10000$ nodes, the choice of $N$ depending on which
aspect must be emphasized. As will be exemplified for some specific
situations, our results remain quite independent of the size of the
network, provided we scale properly the linking probabilities with $N$ for
those networks generated by random algorithms.

\subsection{Clustering coefficient}

\begin{figure}
\begin{center}
\includegraphics*[width=7.6cm,height=5.7cm,angle=0]{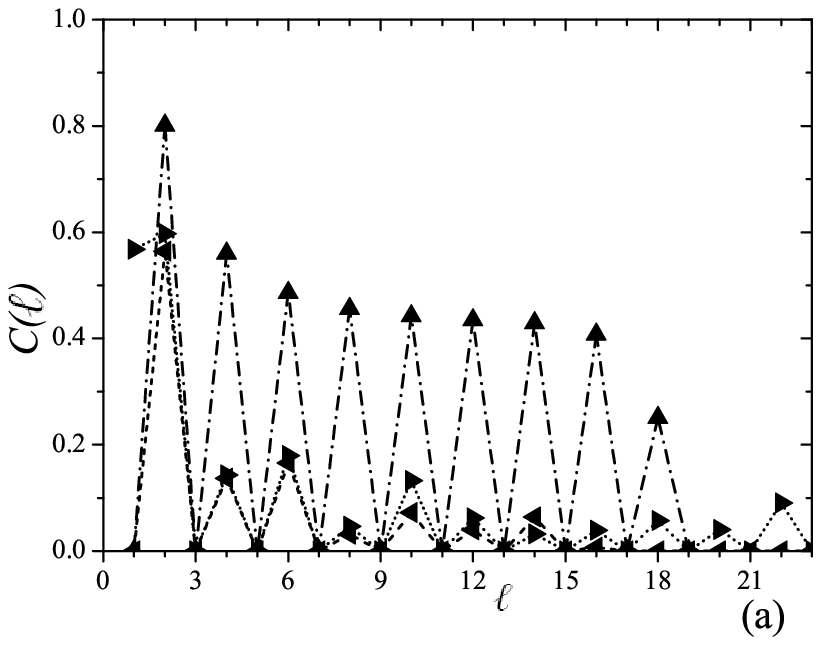}
\includegraphics*[width=8cm,height=6cm,angle=0]{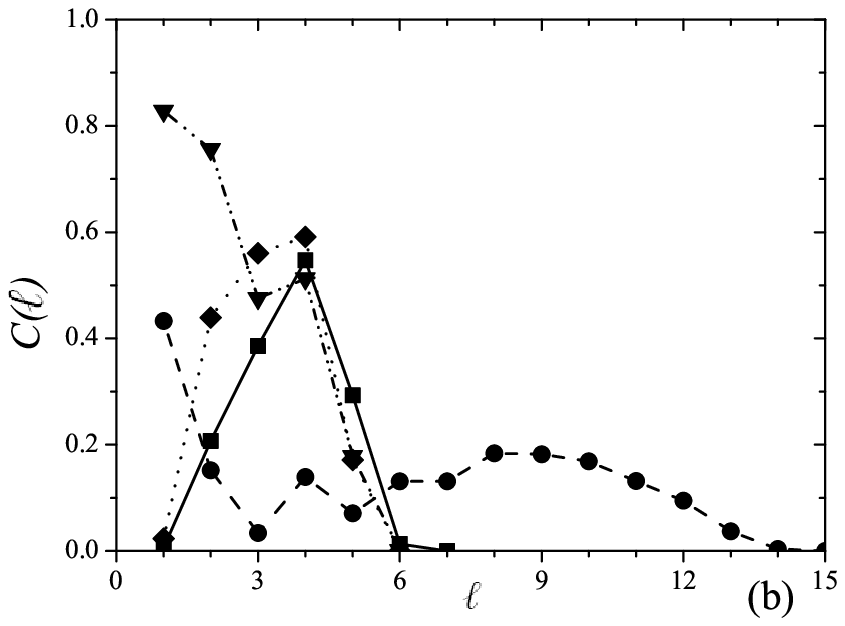}
\end{center}
\caption{Dependence of $C(\ell) \times \ell$ for the seven distinct
networks. Conventions for (symbol, line) introduced here and used in the
rest of this work are as follows: CT (up triangle, dash dot); AN (down
triangle, dash dot dot); DHL (left triangle, short dash); WHL (right
triangle, short dot); SF (diamond, dash); ER (square, solid); SW (circle,
dash). The data were obtained for following values of $N$: (a) 1534 (CT),
684 (DHL), 1564 (WHL). (b) 3283 (AN), 2000 (SF); for ER, $N$=1000 and
$p_a=0.008$; for SW, $N=1000, p_r=0.2$.} \label{fig1}
\end{figure}

Figure 1 shows how $C(\ell)$ behaves for the distinct networks. For the
CT, $C(\ell)$ oscillates between zero and finite values for odd and even
values of $\ell$. This is explained by the fact that, as $\ell$ grows, the
$R(\ell)$ networks assume, alternately, the structures of loop-less trees
and Husimi cactuses, which entail a lot of triangles. This network has
relatively large diameter, so that this behavior is sustained for several
periods. Finite size effects cause a small decrease of the value of
$C(\ell)$, what is enhanced when $\ell$ approaches $D$. Similar behavior
is observed in many other situations, as the regular hyper-cubic lattices
of dimension larger than 1, where triangles appear only for even $\ell$'s.
It is also present in hierarchical structures as DHL. On the other hand,
WHL behaves differently only for $\ell=1$, when $C(1)=0.57$, due to the
presence of triangles. However they are absent for all larger order odd
$\ell>1$, so that an oscillatory pattern between $0$ and non-zero values
of $C(\ell)$ sets in.

A second common pattern for $C(\ell)$ is that of curve with a well defined
maximum, taking finite values only over a finite range. It is found for SF
networks generated according to the standard procedure
\cite{Barabasi_science}, as well as for ER networks when, for instance,
$N=1000$ and connection probability $p=0.008$. For the SF networks, this
patterns emerges because neither the dominant hubs nor the nodes with
lower $k$ are likely to be connected in cliques. However, the hubs induce
many lower $k$'s nodes to form cliques at second and third order
neighborhoods, explaining the sharp increase in the value of $C(\ell)$.
The decrease in the value of $C(\ell)$ as $\ell\rightarrow D$ reflects the
fact that the most part of all pairs of nodes have already been
considered. For ER networks, similar arguments apply.

A third kind of pattern has also been observed for networks that are
characterized by a large value of $C(1)$ as, e.g., the AN and SW
networks. In the first case we start from a large value of
$C(1)=0.828$, following an almost monotonic decrease of $C(\ell)$.
The same is observed for relatively large values of the rewiring
probability $p_r$ in SW networks.

We find important to notice, however, that if the behavior for the
deterministic networks remain essentially the same as $N$ grows, the
same is not observed for SW and ER if we keep $N$ constant and
decrease the values for, respectively, $p_r$ and the linking
probability $p_a$. For both situations, oscillatory pattern between
non-zero values emerges, indicating a similar picture to those of
geometrically constructed networks.

It is worth mentioning that other authors also proposed some
generalized extensions of the notion of the clustering coefficient.
In \cite{Fronczak}, it is investigated the issue if two neighbors of
a given node are far from each other with a chosen distance $x$.
Differently, in \cite{Costa}, the author asks if two nodes, at a
chosen distance $y$ of a given node, are neighbors of each other.
There are also more ways to interpret the concept of neighborhood of
a given node \cite{Caldarelli}. However, as far as we are aware, our
approach, using the notion of neighbors of higher order, goes in
different direction from that of the known literature.

\subsection{Average degree}

For regular lattices, the higher order neighbors of a site are roughly
located on the surface of a hyper-sphere. Their number, which is
equivalent to $\langle k (\ell)\rangle$, grows as $\ell^{dim - 1}$, where
$dim$ denotes the Euclidian dimension of the lattice. For the extreme
situation of an exact CT, where the number of sites on the surface is of
the same order of magnitude as those in the bulk, corresponding to
$dim\rightarrow\infty$. This is reflected  by an exponential increase of
$\langle k(\ell)\rangle$ with $\ell$, as shown in Figure 2a.

\begin{figure}
\begin{center}
\includegraphics*[width=8cm,height=6cm,angle=0]{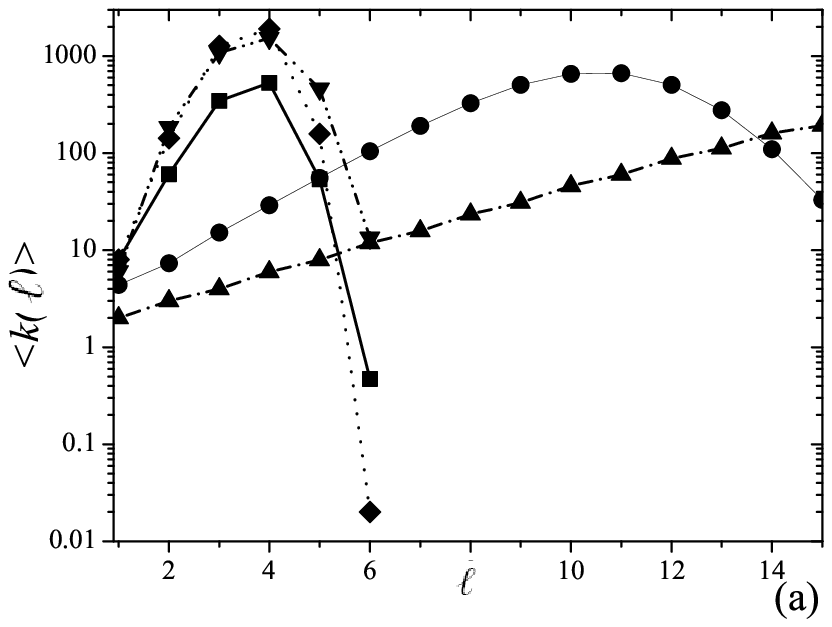}
\includegraphics*[width=8cm,height=6cm,angle=0]{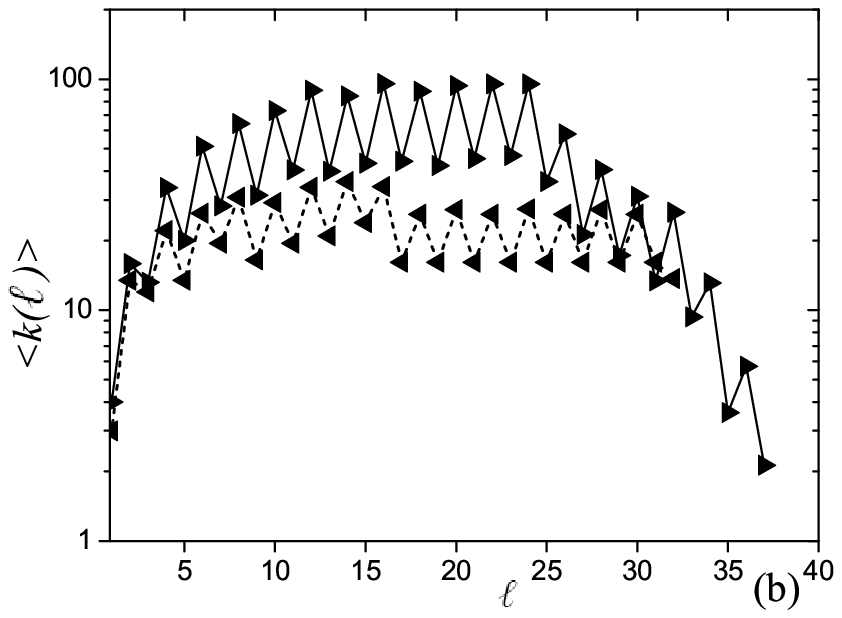}
\end{center}
\caption{Dependence of $\langle k(\ell)\rangle \times \ell$ for
distinct networks. Symbols and lines follow the same convention as
in Figure 1. (a) Values of $N$ are the same used in Figure 1 for CT,
AN, SF and ER, with $p_a=0.008$. For SW, $N=3500$ and $p_r=0.2$. (b)
Same values of $N$ as in Figure 1 for DHL and WHL. } \label{fig2}
\end{figure}

A similar behavior is expected to be found for many other complex
networks, which have a very small diameter. In Figure 2 we
illustrate the behavior of $\langle k (\ell)\rangle$ for the same
set of networks, showing that exponential increase is present for
the majority of them. However, as the network is finite, $\langle k
(\ell)\rangle$ must decrease for sufficiently large value of $\ell$,
making it difficult to assign a precise behavior for the dependence
between $\langle k (\ell)\rangle$ and $\ell$ for these networks. In
Figure 2a we find deviations from the exponential increase already
when $\ell=4$ for some networks. The DHL and WHL (Figure 2b) show a
different pattern, consisting of an oscillating period 2 behavior,
which is caused mainly by the contribution of the nodes introduced
in the last hierarchy. For instance, in the DHL, these nodes are
connected only to two other nodes but, for $\ell = 2$ and $3$, they
can have up to 9 and 4 neighbors. This situation is repeated for
larger values of $\ell$ and also for the WHL.

\subsection{Node distribution probability}

The node distribution $p(k;\ell=1)$ usually displays a bell shaped form
for many kinds of networks found in nature, as well as those generated by
the ER (Figure 3a) and SW (not shown) algorithms. Social and natural SF
networks, the features of which we reproduce by the preferential
attachment growth algorithm, are representative of another possible
pattern, characterized by $p(k;1)\sim k^{-\gamma}$. For the purpose of a
clearer picture, there we draw $
\underline{P}(k;\ell)=\frac{1}{k}\int_{k}^{\infty}p(k';\ell)dk'$ for the
usual SF, reproducing the value $\gamma\simeq 3$, as shown in Figure 3b
for $\ell=1$. In our investigations, we also find similar behavior for the
AN, DHL and WHL networks, as shown in the points for $p(k)$ in the inset
of this figure. The CT is a trivial case where $p(k;1)$ reduces to a
single point.

\begin{figure}
\begin{center}
\includegraphics*[width=8cm,height=6cm,angle=0]{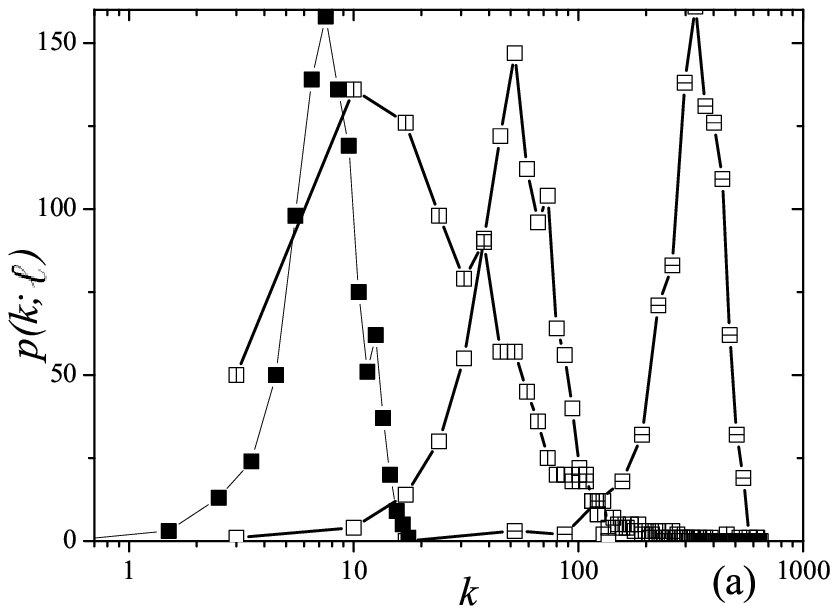}
\includegraphics*[width=8cm,height=6cm,angle=0]{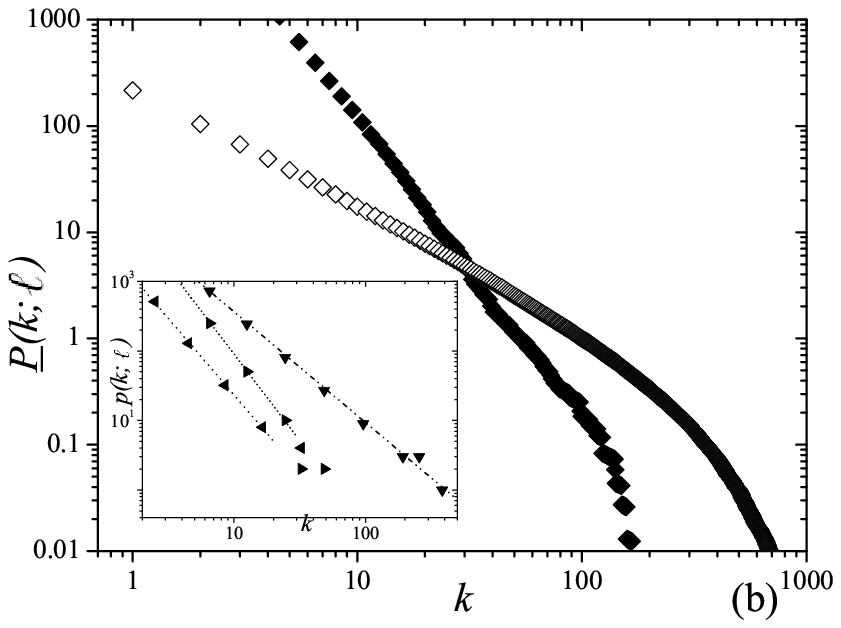}
\end{center}
\caption{Dependence of the probability distribution $p(k;\ell)$ and
cumulative probability distribution $\underline{P}(k;\ell) \times k$
for distinct networks. (a) Curves for ER with same values of $N$ and
$p_a$ used in Figures 1 and 2, when $\ell=1$ (solid), 2 (hollow), 3
(horizontal dash) and 5 (vertical dash) squares. (b) In the large
panel, $\underline{P}(k;\ell)$ follows power law for SF (when
$\ell=1$ (solid) and $\ell=5=D-1$ (hollow)); in the inset, points
indicate power law decay for $p(k;\ell=1)$ in AN, DHL and WHL.}
\label{fig3}
\end{figure}

For $\ell>1$, our results show that the bell shaped pattern for
$p(k;\ell)$ is reproduced when $\ell>1$ for the ER and SW networks (see
Figure 3a). As shown before, the average node degree increases with
$\ell$, so that it is natural that the curves for $p(k;\ell)$ are shifted
towards larger values of $k$. The situation changes only for sufficient
large $\ell \simeq D$, when finite size effects become relevant. Then, the
majority of nodes have already reached their most distant neighbors while
others miss only few connections. In this situation, the few non zero
contributions to $p(k;\ell\simeq D)$ come for values of $k\sim 1$, so that
the peak of the curve is shifted to the region close to the origin, as
shown, in Figure 3a, for $\ell=5$ .

For the SF networks, the regions where the higher order $p(k;\ell)$'s
receive significant contributions are also pushed to large values of
$k(\ell)$. In general, the distributions loose the power law behavior,
assuming distinct forms as $\ell$ increases. Exceptions are provided,
e.g., in the example shown for the cumulative distribution
$\underline{P}(k;\ell)$ in the large panel of Figure 3b. There we find a
very interesting return to a power law distribution in the region of low
values of $k$ when $\ell=D-1$, with an exponent $\gamma\simeq1.2$.  To
conclude this analysis, we remark that both DHL and WHL fail to display
any noticeable alignment of points in any of their higher order
neighborhood.

\subsection{Hierarchical property}

There are several distinct concepts of hierarchical organization,
some of them stemming from the network framework, others from
geometrical constructions, self similar fractal sets, etc. In this
work, apart from using the word "hierarchical" to denote DHL and
WHL, we refer to the concept introduced in \cite{Ravasz},
according to which a network has the hierarchical property if
$C_i(1)$, the clustering coefficient of an individual node $i$,
when are drawn as function of the individual node degree $k_i(1)$,
shows a power law decrease. Much as observed with the analysis of
the node distribution $p(k;\ell)$, evidences of hierarchical
character for $R(\ell>1$), networks are rare, and are only found
when $R(1)$ has already such character.

\begin{figure}
\begin{center}
\includegraphics*[width=7.2cm,height=5.4cm,angle=0]{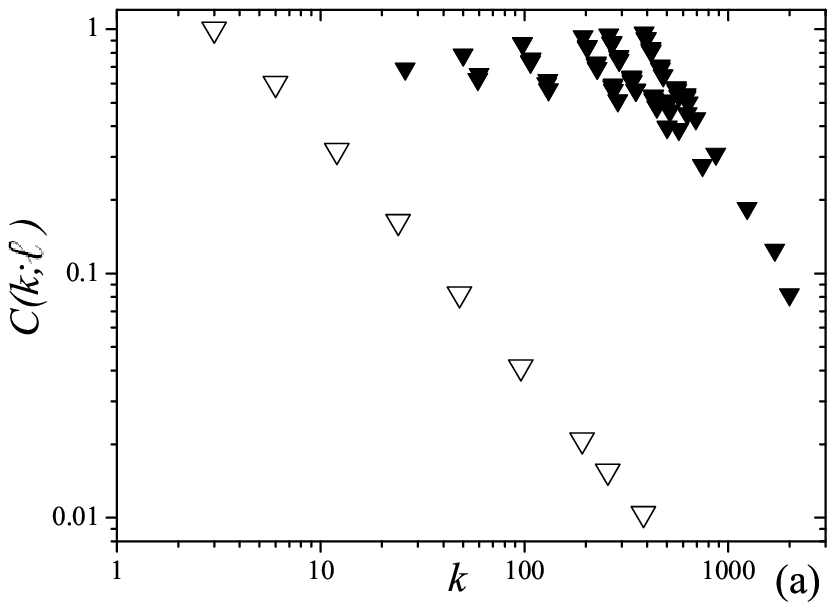}
\includegraphics*[width=8cm,height=6cm,angle=0]{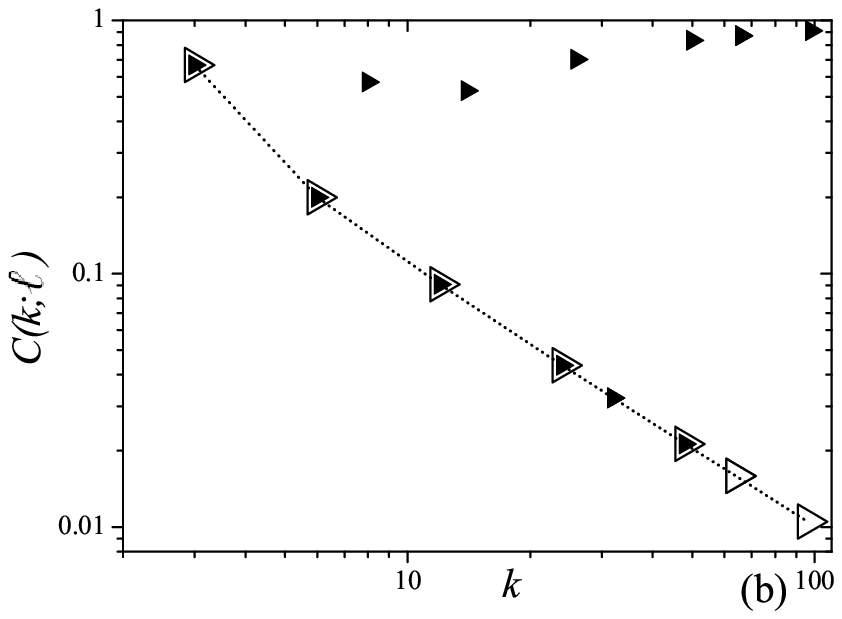}
\end{center}
\caption{Data for $C(k;\ell) \times k$ indicating hierarchical
arrangement of nodes for AN (a) and WHL (b), respectively with
$N=3283$ and 7814 nodes. Data are drawn for $\ell=1$ (hollow) and
$\ell=2$ (solid). } \label{fig4}
\end{figure}

For the distinct networks we work with, indications of power law
behavior for $\ell\geq 1$ are found for the AN and WHL networks, as
illustrated in Figure 4a and 4b. Both networks show $C(k;1)\sim
k^{-1}$ asymptotic dependence. On the other hand, DHL does not show
this property since, as discussed before, $c_i(\ell=1)\equiv0,
\forall i$.

We find slight evidences, for $\ell>1$, of power law dependence in some
subsets of points. In the particular case $\ell=2$, for the AN, several
points are aligned, seemingly building short patches that obey power law
decrease. This very peculiar distribution of points is recurrent for all
values of $N$ that correspond to a full generation of the construction of
AN.

For WHL, several points of the curves for $C(k;1)$ and $C(k;2)$
coincide exactly, so that a subset of points of $C(k;2)\times k$
still follows a power law decay. This is exemplary shown in Figure
4b, for the 7-th generation (7814 nodes). To explain the presence of
this coincident points, we must take into account that, on
increasing the generation of WHL, we add points to the region of
large values of $k$ of the $C(k;1)$ curve, while conserving all but
one point of the previous generation. In Figure 4b, two newly added
points, $k=64$ and 96, are absent in the curve for the 6-th
generation. They correspond to nodes with very large degree, e.g. at
the root sites or the intermediate position, where the main bridge
connecting two branches is placed. Now, when we look for pairs of
second neighbors of the 7-th generation, we find that part of them
coincides exactly with first neighbors in the 6-th generation, so
that the $C(k;2)\times k$ curve of the 7-th generation, for this
subset, falls on the top the $C(k;1)$ curve of the 6-th generation.
The other subset, which contributes to points that fall off the
straight line, is formed by pairs of second neighbors that do not
correspond to any first neighbors of the previous generation. For
larger values of $\ell$, we should still observe this truncated
pattern only for the even values of $\ell$. Indeed, as already
discussed before, $C(\ell)=0$ for odd $\ell$'s.

\subsection{Assortativity degree coefficient}

Several assortativity properties can be assigned to a network
\cite{Dorogovtsev}. Each of them is quantified by a corresponding
coefficient $A$, which indicates whether the pairs of nodes
directly connected by a link are more likely to behave alike
($A>0$) or dislike ($A<0$). The assortativity degree coefficient
\cite{Newman_assortativity}, which takes into account the average
degree of the nearest neighbors of a node of degree $k$
represented by $k_{nn}(k)$, probes the degrees of the nodes at
each side of a link. Here, we denote by $A(\ell)$ the coefficients
that quantify the degree assortativity for the corresponding
neighborhoods $O(\ell)$. Each $A(\ell)$ measures whether pairs of
nodes, that are $\ell$ steps apart, are likely to be
$\ell$-connected to other nodes that have the same $\ell$ degree
$k(\ell)$.

\begin{figure}
\begin{center}
\includegraphics*[width=8cm,height=6cm,angle=0]{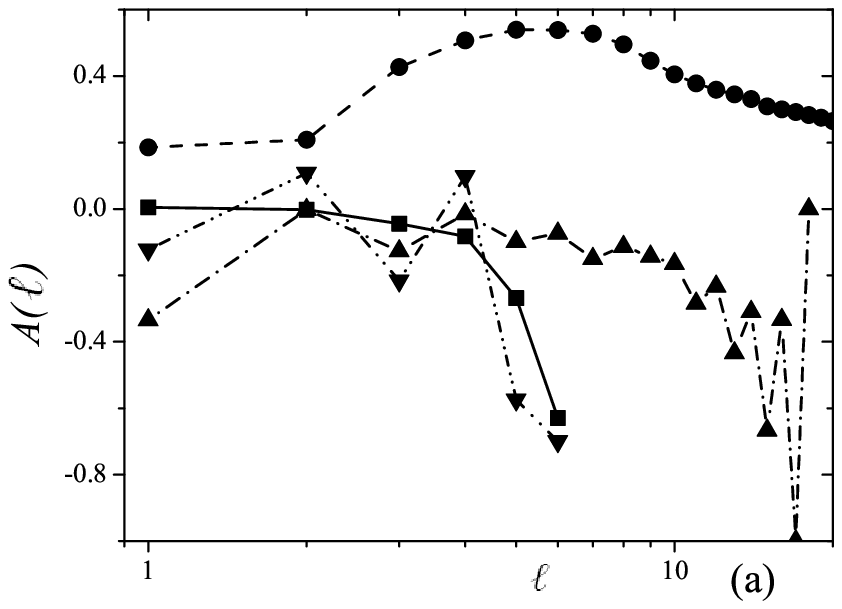}
\includegraphics*[width=8cm,height=6cm,angle=0]{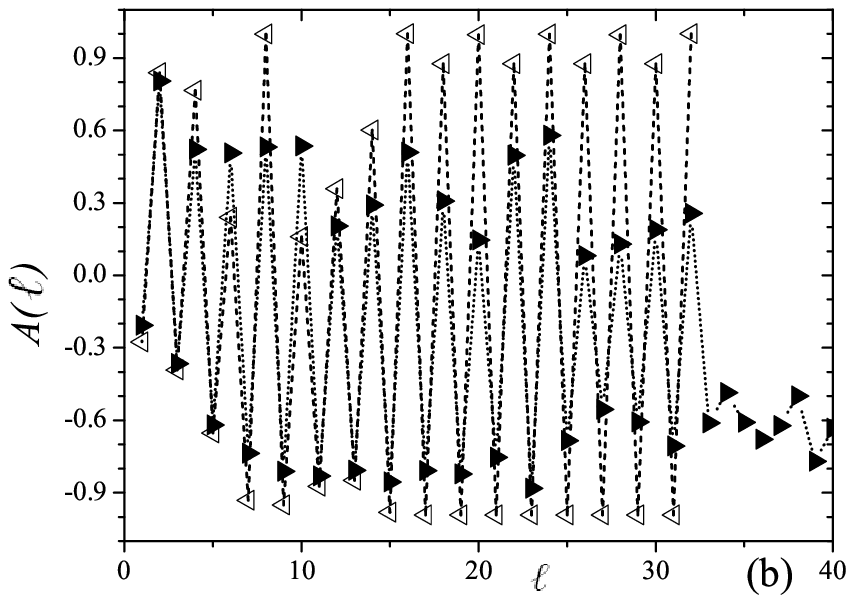}
\end{center}
\caption{Dependence of the degree assortativity coefficient $A(\ell)$ with
$\ell$. (a) Examples of assortative behavior for SW network, with $N=3500$
and $p_r=0.2$; neutral behavior for ER, with same parameter values as in
Figure 1; oscillatory character for AN (N=3283); dissortative behavior for
CT (N=1534). (b) Results for DHL and WHL with the same values of $N$ as in
Figure 1, indicate $\ell=2$ periodic oscillations from dissortative to
assortative behavior .}\label{fig5}
\end{figure}

The results in Figure 5a-b display several kinds of patterns, indicating
that the behavior of $A(\ell)\times \ell$ is very sensitive to the type of
network. Figure 5a shows that SW networks have usual positive
assortativity $A(\ell=1)\simeq 0.2$. This value results from the
contribution of the large majority of sites that are connected to their
original neighbors in an ordered structure. On increasing $\ell$,
$A(\ell)$ remains positive for a large $\ell$ interval. After this phase,
$A(\ell)$ goes through a steep descent to negative values, where it
remains until $\ell=D$. This change reveals that the large $\ell$
neighborhoods completely loose the local character and nodes are
overwhelmingly $\ell$-connected to nodes with distinct $\ell$-degree.

The results for a finite CT, in the same panel, are strongly biased
by surface effects. In an infinite tree, all nodes have the same
degree, so that $A(\ell=1)=1$. However, for any finite tree, a
dissortative character is observed, which can be explained as
follows: (i) The number of sites added in the last generation has
the same order of magnitude of the existing sites; (ii) All of them
have a distinct number of neighbors as those added before, the same
happening to the nodes that are connected to them. In the evaluation
of $A(1)$, nodes with distinct degree on the end of all newly
introduced links, contribute negatively, so that these contributions
lead to a dissortative character to CT. The same occurs for larger
values of $\ell$, so that all $A(\ell)$ deviate strongly from the
constant value 1 that they assume in an infinite tree. However, the
constant value 1 can be recovered, in a finite sized network, if one
neglects, successively, the effect of the $\ell$-th lately added
nodes.

For the ER networks, also illustrated in Figure 5a, pure randomness shows
neutral behavior, hence $A(1)=0$. Numerical simulations reproduce this
result, which should be valid for several values of $\ell$. However, for
large $\ell$, finite size effects end up by driving $A(\ell)$ to the
negative region.

With respect to geometrically grown networks, oscillations between
dissortative to assortative character is a common feature for AN, DHL and
WHL. In the first situation (Figure 5a), the amplitude of variation of
$A(\ell)$ is not so large and, due to the very small value of $D$, short
lived. On the other hand, for DHL we have very large variations limited
only by the extremal values $\pm 1$ (see Figure 5b). This very peculiar
behavior may be explained by noting that, for $\ell=1$, no node has
neighbors with the same degree as itself. On the other hand, when
$\ell=2$, the number of nodes with second order neighbors having the same
degree is very large. Thus, oscillations set it, changing the assortative
character at each step. For the WHL, very large oscillations are also
observed, although they do not reach the extremal values as for DHL.
Indeed, in this situation, the presence of the cross bonds lead to a large
heterogeneity in the degree of the nodes, what causes a decrease in the
amplitudes of oscillations.

\subsection{Fractal dimension}

As discussed in Section II, we can make use of the results
obtained in the evaluation of the higher order neighborhoods to
obtain the network fractal dimension $d_F$ as proposed in .
Results for the distinct networks are summarized in Figure 6.
Confirming results reported in \cite{Havlin}, it shows that, for
many networks, this definition leads to a power law dependence
between the quantities $L(\ell)$ and $\ell$. Although our results
support the definition for $d_F$, it also confirms that finite
size effects and very short diameter raises intrinsic difficulties
to its evaluation for many networks.

\begin{figure}
\begin{center}
\includegraphics*[width=8cm,height=6cm,angle=0]{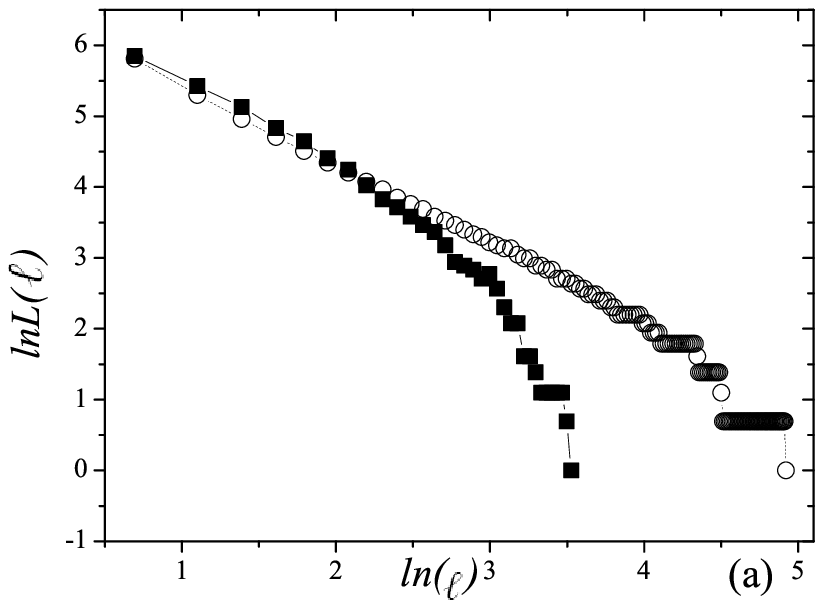}
\includegraphics*[width=8cm,height=6cm,angle=0]{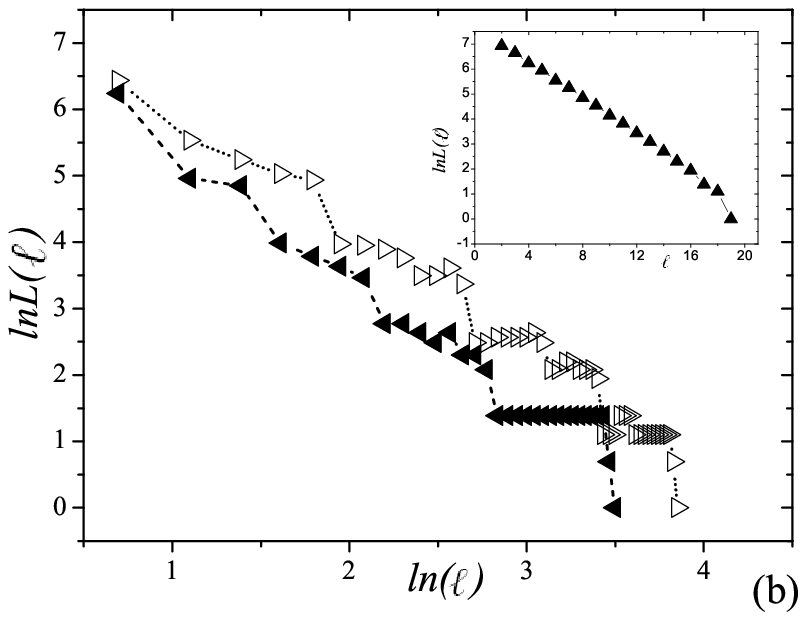}
\end{center}
\caption{Scaling regime between $L(\ell) \times \ell$. (a) Results for SW
($N=1000,p_r=0.01$) and ER ($N=580,p_a=0.002$), where scaling is quite
evident. (b) In the large panel, results for DHL and WHL (with same $N$ as
in Figure 1). In the inset, results for CT (N=1534), with exponential
dependence indicating infinite Euclidian and fractal dimensions.}
\label{fig6}
\end{figure}

So, we see in Figure 6a that, for SW networks, well defined scaling
region can be found for very small values of $p_r$, when $D$ becomes
large enough. In Figure 6b, DHL and WHL also show scaling behavior,
although the decrease of $L(\ell)$ proceeds through a series of
large size steps. The slope of both curves are similar ($\sim
1.82$), but differ from the values obtained for the usual fractal
dimension \cite{Melrose}. We observe, in the inset of Figure 6a,
that the CT definitely follows an exponential rather than power law
decay. This corresponds to an infinite value for $d_F$, what is in
agreement with the fact that it is not possible to associate a
finite dimension to the tree, where the number of sites in the bulk
and surface have the same order of magnitude. These results confirm
that the definition of $d_F$ is sound, capturing the essential
features of quite distinct networks but, as in the case of DHL,
values for $d_F$ may differ from those obtained from other
definitions.

For AN and SF networks, where the only free parameter is the number of
nodes, a roughly  linear alignment of points, like those in Figure 6a, is
also identified, whenever over only a short range of $\ell$, limited by
the small values of $D$. For ER networks, this is also the typical
behavior, unless $p_a$ is very small, as illustrated in Figure 6a. There,
a very small value of $p_a$ warrants a fairly well defined scaling region
between $L(\ell)$ and $\ell$.

On closing this section, we would like to recall that the investigation of
higher order neighborhoods for some well known networks opens the door to
a large number of results that have not been taken into account sofar. In
all subsections we were able to reproduce known results for the network
indices when $\ell=1$ and, at the same time, present some very peculiar
features for the same parameters when evaluated at larger values of
$\ell$. In this first investigation we make the the decision to draw
definite conclusions only from the most relevant situations, but we are
aware that, within the large amount of information we obtained, many
secondary aspects of our investigation have not been fully discussed.
Nevertheless, we are sure to have shown that this framework offers, many
perspectives to the understanding and characterization of the larger
structures in a complex network. After exploring the $M(\ell)$'s matrices
on their own, in the next section we will push forward a new investigation
based on the assembly of all of them, in a single matrix
$\mathbf{\widehat{M}}$. Actually, we explore the possibility of defining a
quantitative measure distance between networks with the same number of
nodes, based on the properties of their higher order neighborhoods.

\section{Distance concepts in network neighborhood space}

In the last Section, we developed a systematic procedure to obtain
more details on the structure of networks by evaluating a set of
known indices for their higher order neighborhoods.  Now we want to
explore the information of the neighborhood structure contained in
$\mathbf{\widehat{M}}$ to another difficult challenge, namely
\textit{comparing} and \textit{measuring} distance among networks.
Addressing this task requires the attention to several issues
regarding definitions and procedures, and can not be completely
covered in a single step. Here, we want to tackle some aspects of
this general problem, suggesting measures and methods to test
whether they are particularly suitable to allow for such comparison.

Actually, this issue constitutes an extension of a classical
problem within the framework of graph theory, namely, once we are
given two graphs, to decide whether they are isomorphic or not.
 It is in fact a hard issue: the isomorphism problem for
graphs is one of the non-polynomial (NP) questions, which remains still
unknown whether it is NP-complete or not \cite{Toran}. It has been
followed mostly by mathematicians and computer scientists, who have
developed several methods, algorithms and computer softwares that address
this question, among these, one very fast and known is NAUTY \cite{McKay},
which classifies nodes according information about their immediate
neighborhood; however there are also proposals \cite{Miyazaki} to improve
this method considering neighbors of higher distance. A positive answer to
the isomorphism question, which is very rare due to the multitude of
distinct networks, completely solves the network comparison problem.
However, a negative one does not advance much in providing a measure of
how close the two networks (or graphs) are. The identification of all
neighborhoods of a network and of the neighborhood matrix
$\mathbf{\widehat{M}}$ opens perspectives for addressing the question of
comparison between networks.

\begin{figure}
\begin{center}
\includegraphics*[width=5.cm,height=5.6cm,angle=0]{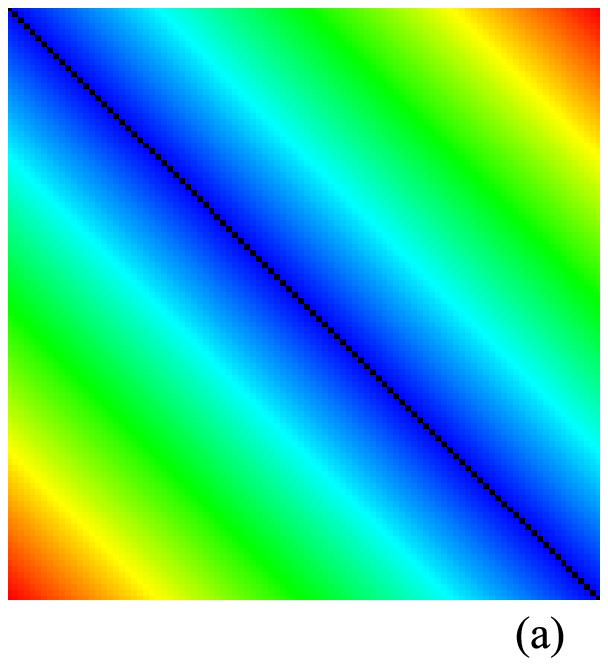}
\includegraphics*[width=8.cm,height=5.8cm,angle=0]{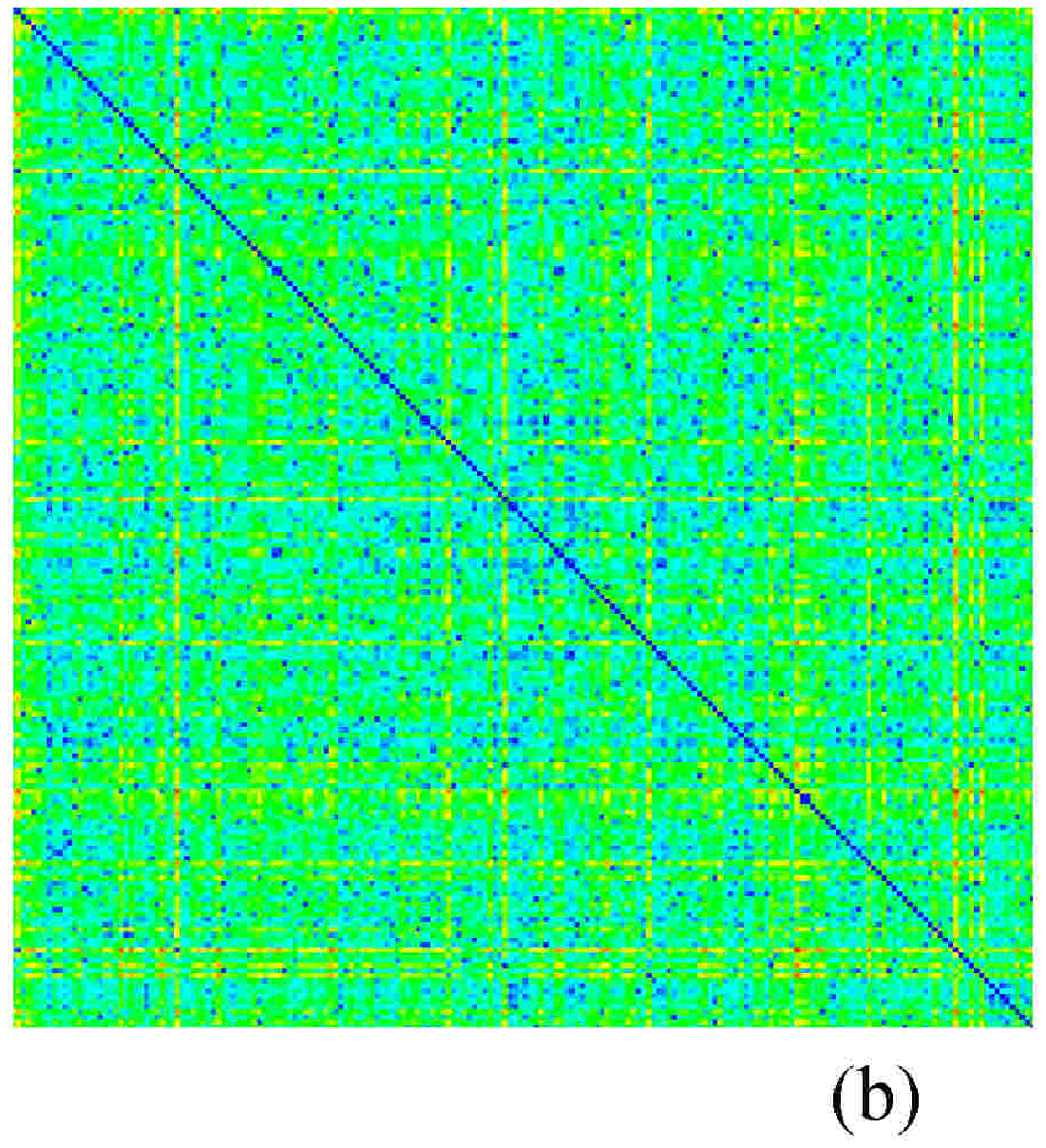}
\includegraphics*[width=8.cm,height=5.8cm,angle=0]{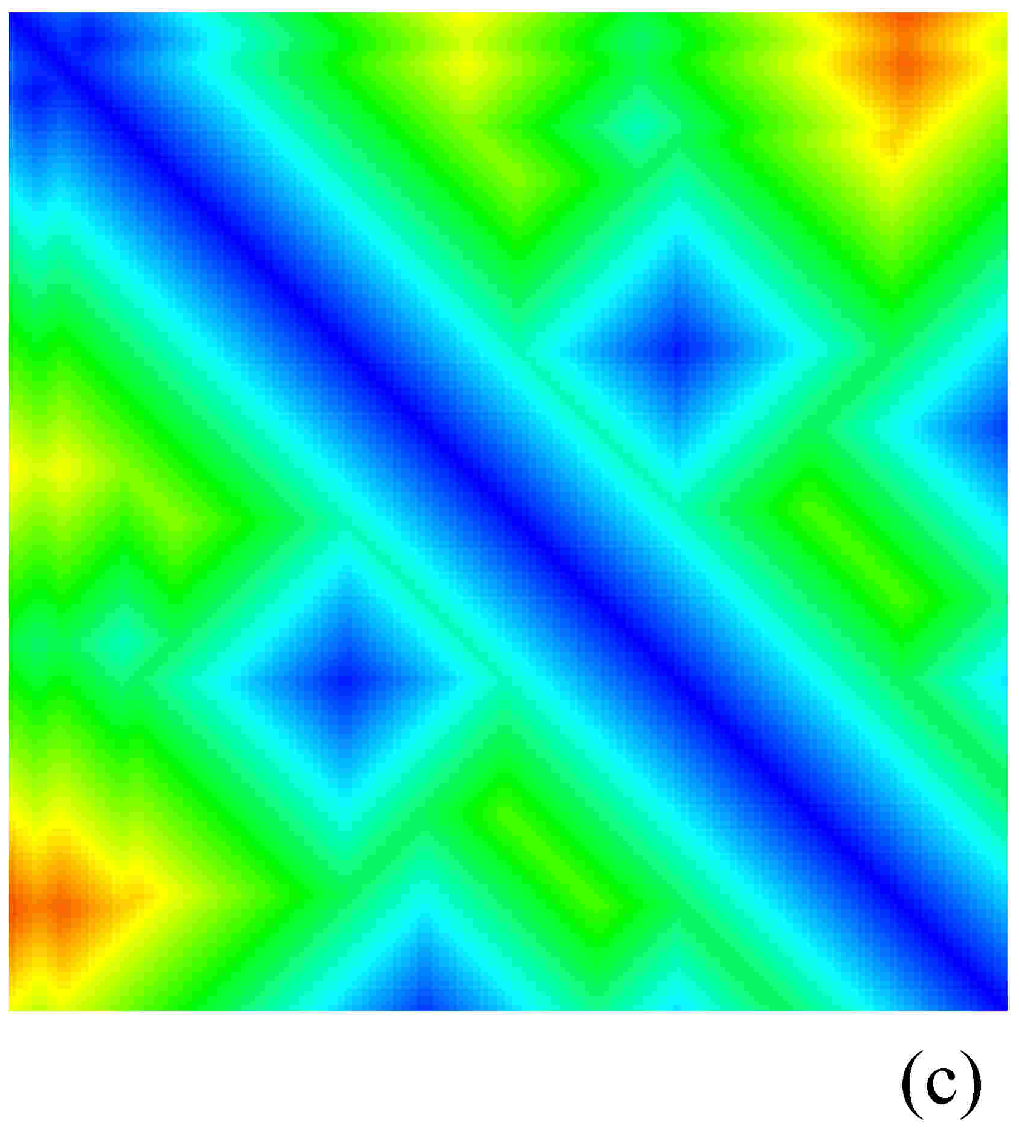}
\includegraphics*[width=8.cm,height=5.8cm,angle=0]{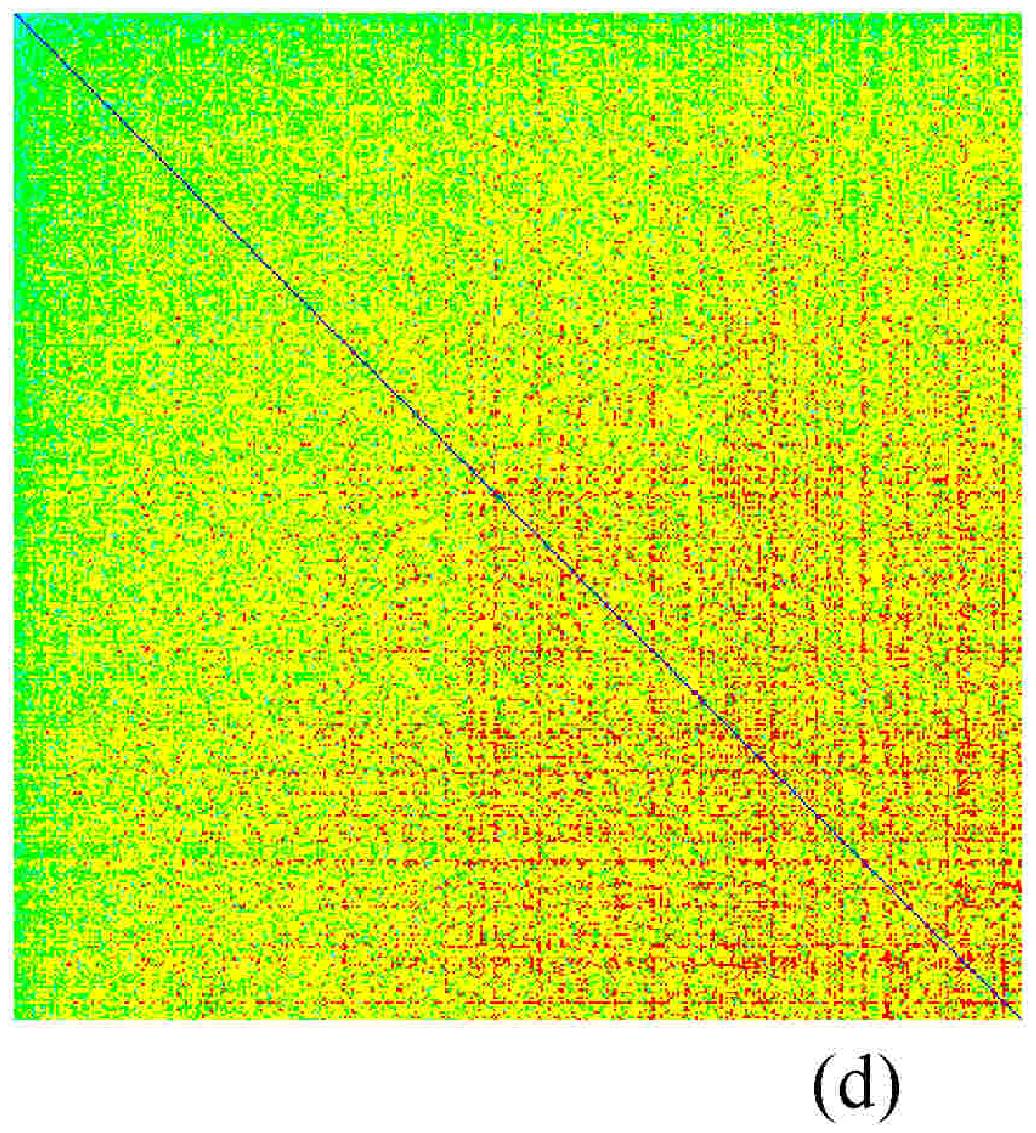}
\end{center}
\caption{Color plots of $\mathbf{\widehat{M}}$ for networks LC (a), ER
(b), SW (c) and SF (d). Color runs from blue when $\ell=1$ through red,
for $\ell=D$. Number of color levels is equal to $D$. Values of ($N$,$D$)
for (a)-(d) are given by, respectively, $(100,50), (582,33), (1000,136),
(1000,6).$ In (b), $p_a=0.002$ and in (c) $p_r=0.01$.} \label{fig7}
\end{figure}

First let us note that, as stated before, color or gray code plots
offers the possibility of visualizing the full network neighborhood
structure, as exemplified in Figure 7a-d. There, we show some
examples of the neighborhood plots, including in (a) that for the
open end linear chain (LC) with links between nearest and
next-nearest neighbors. Apart from the open end detail, LC coincides
with the used SW networks with $p_a=0$. The diameter of LC is
$D=[N/2]$, resulting in a smoothly varying colors from the diagonal
to the opposite matrix corners. The other three examples illustrate
the situation for some other typical (ER, SW and SF) networks we
investigate in this work, making it is possible to recognize,
respectively: b) full randomness of colors in ER network; c) the
emergence of a few islands (in blue) due to the long range links
added over the LC structure on the top of which SW is built; d) a
color gradient towards the upper-left corner of
$\mathbf{\widehat{M}}$, indicating that the low numbered sites,
which have been introduced in early stages of the construction of
the SF network, are at a much smaller number of steps apart from the
other sites.

It is important to call the attention that the representation of a network
by its color/gray plot of $\mathbf{\widehat{M}}$ depends on the particular
numbering for the nodes that is being used. For the examples shown in
Figure 7, the numbering follows the most natural way from the construction
algorithm for the three networks, i.e., by the order in which they are
introduced in the network. On the other hand, we illustrate the dependence
of the color patterns on the used numbering in Figure 8, with two images
for the same AN. The first one is based on an specific numbering scheme
used before \cite{Andrade_spectrum}), while the second one results from
numbering the nodes according to their decreasing degree. We notice in
Figure 8b that the structure for large values of $\ell$, represented by
red spots, is preserved by the renumbering, but moved for the region of
larger values of $(i,j)$.

\begin{figure}
\begin{center}
\includegraphics*[width=6.8cm,height=6.8cm,angle=0]{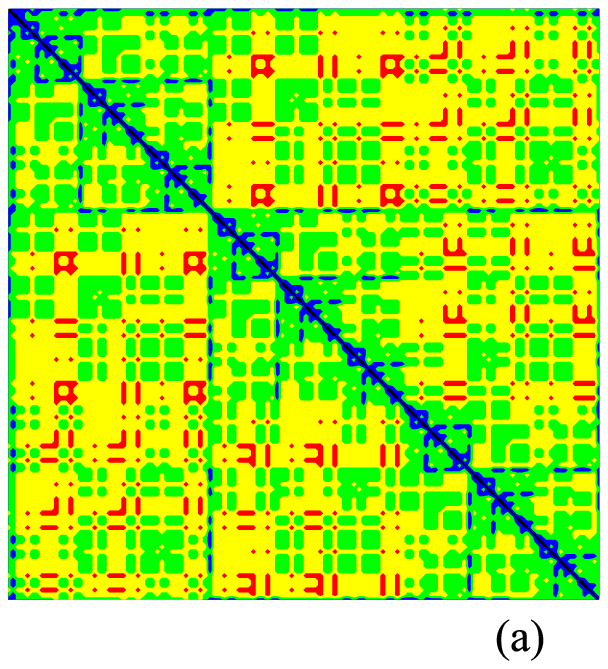}
\includegraphics*[width=6.8cm,height=6.8cm,angle=0]{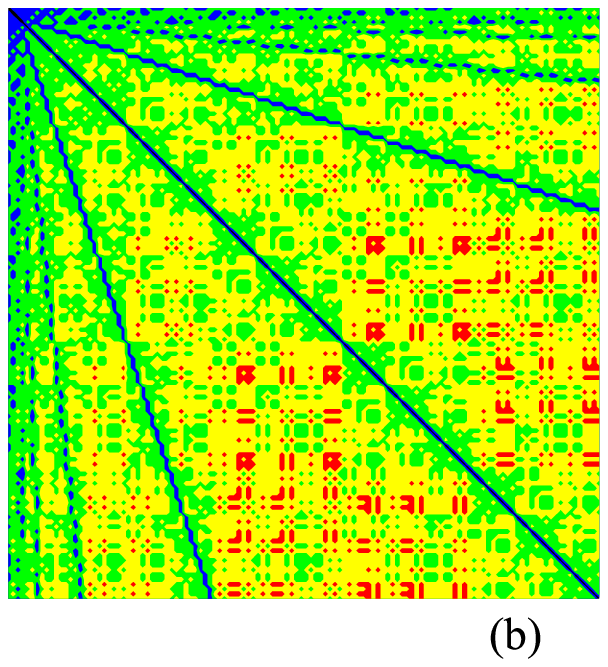}
\end{center}
\caption{Color representation of the $\mathbf{\widehat{M}}$ for AN, when
$N=124$, $D=4$, with two distinct numberings.  (a) Numbering defined in
\cite{Andrade_spectrum}. (b) Starting from numbering in (a), and ordering
numbers according to decreasing order of the node degree. } \label{fig8}
\end{figure}

These images suggest, in a quite natural way, to compare two
networks with the help of the corresponding
$\mathbf{\widehat{M}}$'s, both of which, with exception of the
diagonal, have only nonzero matrix elements. We restrict our
comparison between networks that have exactly the same number of
nodes $N$. Thus, the basic principle behind the methodology  is to
measure a \textit{Euclidian-like distance}
$\mathfrak{D}(\alpha,\beta)$, between two networks $\alpha$ and
$\beta$, by summing over the $positive$ differences between matrix
elements of the two corresponding neighborhood matrices
$\widehat{M_{\alpha}}$ and $\widehat{M_{\beta}}$,
\begin{equation}\label{eq5}
\mathfrak{D}^2(\alpha,\beta) =
\frac{1}{N(N-1)}\sum_{i,j=1}^{N}((\widehat{M_\alpha})_{i,j}-(\widehat{M_\beta})_{i,j})^2.
\end{equation}

Note that we consider the distance per matrix element, and that
other definitions, based e.g., on the absolute values between the
matrix elements, can be used within the proposed framework as well.
We would like to stress that networks $\alpha$ and $\beta$ can only
be identical if they also share the same value of $D$ and the same
set of links $L(\ell), \ell=1,...,D$. Otherwise,
$\mathfrak{D}(\alpha,\beta)$ can never vanish. We note also that
$\mathfrak{D}(\alpha,\beta)$ could hardly have the meaning of an
actual distance if defined only on the basis of the information in
the adjacency matrices $M_{\alpha}(1)$ and $M_{\beta}(1)$. Indeed,
the fact that these are prone of zero elements would rend much less
precise the results based on a similar scheme as the one we propose.
In other words, $\widehat{M_\beta}$ is able to set up a much more
precise measure of distance than $M(\ell=1)$.

One may ask whether it is possible to extend this procedure in order
to compare networks $\alpha$ and $\beta$ with distinct diameters
$D_\alpha \neq D_\beta$. In such situation, the networks are never
isomorphic, but one can also ask how far apart they are, in the
sense that pairs of nodes that belong to  $O_\alpha(\ell=1)$ are
close to those $O_\beta(\ell=1)$, while those in
$O_\alpha(\ell=D_\alpha)$ are close to those
$O_\beta(\ell=D_\beta)$. Of course, pairs of nodes with intermediary
values of $\ell$ in both networks should also come closer to one
another. We find out that, to this purpose, it becomes necessary to
redefine $\mathfrak{D}(\alpha,\beta)$.

Indeed, equation (\ref{eq5}) does not properly take into account the
distinct contributions in order to warrant an interpretation of
distance as discussed before: pairs of sites in $O(\ell=1)$ can
indeed vanish their contribution to $\mathfrak{D}(\alpha,\beta)$,
but those at $O(\ell=\max(D_\alpha,D_\beta))$ will never succeed
doing the same. Therefore, we introduce a new distance
$\mathfrak{d}(\alpha,\beta)$ according to
\begin{equation}\label{eq6}
\mathfrak{d}^2(\alpha,\beta) = \frac{1}{N(N-1)}\sum_{i,j=1}^{N} \left [
\frac{(\widehat{M_\alpha})_{i,j}}{D_\alpha}-\frac{(\widehat{M_\beta})_{i,j}}{D_\beta}
\right ] ^2.
\end{equation}
We clearly see that, on normalizing the terms of
$\mathfrak{d}(\alpha,\beta)$ to the $[0,1]$ interval, it differs
from $\mathfrak{D}(\alpha,\beta)$ only by a constant factor when
$D_\alpha=D_\beta$. For the more general situations $D_\alpha\neq
D_\beta$, this definition avoids the large contributions coming from
quite distinct values of $\widehat{M}$ to
$\mathfrak{D}(\alpha,\beta)$, and accomplishes our intention to
properly compare pairs of nodes in the whole $\ell$ interval.

If we proceed within graph theory and find that the networks
$\alpha$ and $\beta$ are isomorphic, this indicates that they are
essentially the same. Thus, they differ only by the choice made for
numbering the sites, but share the same topology, which depends only
on the way the links are distributed among pairs of nodes. On the
other hand, if we find that, for this specific choice of isomorphic
networks, $\mathfrak{d}(\alpha,\beta)\neq 0$, we must conclude that
their nodes are numbered differently, this result just reflecting
the fact that $\mathbf{\widehat{M}}$ depends on the way the nodes
are numbered. If we enumerate both networks according to one same
rule, we obtain $\mathfrak{d}(\alpha,\beta)= 0$

In a general situation, given two arbitrary networks, it is expected that
$\mathfrak{d}(\alpha,\beta)\neq 0$. However, since
$\mathfrak{d}(\alpha,\beta)$ depends on the used node numbering, it is
possible to scan over the space of $N!$ distinct numberings for one of the
lattices, say $R_\beta$, in order to identify, decrease or even eliminate
nonzero contributions to $\mathfrak{d}(\alpha,\beta)$ stemming from the
node numbering. Differences in $\mathfrak{d}(\alpha,\beta)$ resulting from
distinct topologies of the networks are essential, and can not be removed
by the renumbering procedure.

To accomplish this purpose, after being sure that
$\mathfrak{d}(\alpha,\beta)\neq 0$, we use a Monte Carlo (MC) algorithm in
order to find a better numbering for the $\beta$ network, in the sense
that the value of $\mathfrak{d}(\alpha,\beta)$ is reduced, while holding
fixed the original numbering for the $\alpha$ network. In each MC step,
the number of two randomly chosen sites (say $s$ and $t$) of $R_\beta$ are
exchanged if $\mathfrak{d}'(\alpha,\beta)<\mathfrak{d}(\alpha,\beta)$,
where $\mathfrak{d}'(\alpha,\beta)$ represents the new distance, evaluated
by taking into account renumbering lines and columns of
$\widehat{M_\beta}$ by letting $s\rightleftarrows t$. If
$\mathfrak{d}'(\alpha,\beta)>\mathfrak{d}(\alpha,\beta)$, renumbering
occurs with probability $\sim \exp(-(\mathfrak{d}'-\mathfrak{d})/T)$,
where $T$ is the usual Monte Carlo "temperature".

With similar results, we have also succeeded in reducing
$\mathfrak{d}(\alpha,\beta)$ with a deterministic procedure. Here, we
first scrutinize the terms in (\ref{eq6}), looking for the line in matrix
$\widehat{M_\beta}$ (say $s$) with the largest contribution to
$\mathfrak{d}$. After this step, we identify the line $t$ in
$\widehat{M_\beta}$ which, moved to the position $s$, minimizes the value
of  $\mathfrak{d}$, and exchange the corresponding lines and columns
$s\rightleftarrows t$. If none of the lines produces a smaller value for
$\mathfrak{d}(\alpha,\beta)$ than $s$, we take the line with next largest
contribution to $\mathfrak{d}(\alpha,\beta)$, and so on.

We use these two procedures to find (local) minima of
$\mathfrak{d}^{min}(\alpha,\beta)$, which may eventually correspond to the
absolute minimum. Whether this last possibility has been met can only be
decided after several runs of the randomized MC algorithm, or if we
succeed finding $\mathfrak{d}^{min}(\alpha,\beta)=0$ by either of the
renumbering procedures.  Of course we also expect that
$\mathfrak{d}^{min}(\alpha,\beta)=\mathfrak{d}^{min}(\beta,\alpha)$, but
we should keep track that, in both renumbering procedures, the first
network keeps the original numbering. Therefore, after performing the
minimization to find $\mathfrak{d}^{min}(\alpha,\beta)$, we obtain the
numbering for $R_\beta$ that brings it as close as possible to that of the
original numbering of $R_\alpha$, the corresponding matrix of which is
denoted as $\mathbf{\widehat{M}}_{\beta \rightarrow \alpha}$. Loosely
speaking, it represents, in the space of all possible networks, a kind of
projection of the $R_\beta$ onto the direction defined by $R_\alpha$. We
will use this concept in the discussion of our results in the next
section.

\section{Evaluation of network neighborhood distances}

In order to validate the minimization methodologies described before, we
compared several pairs of identical networks. Each pair was obtained from
one given $\widehat{M_\alpha}$, while $\widehat{M_\beta}$ is obtained by
randomly  shuffling pairs of lines and columns. This procedure ensures
that both networks share the same topology, differing only by the label
assigned to each node. Thus, the distance
$\mathfrak{d}^{min}(\alpha,\beta)$ is expected vanish if the minimization
strategies work properly. Both algorithms proved to be reliable, being
able to reach $\mathfrak{d}(\alpha,\beta)=0$ for a very large set of
number-randomized network pairs.

\begin{figure}
\begin{center}
\includegraphics*[width=8.cm,height=5.8cm,angle=0]{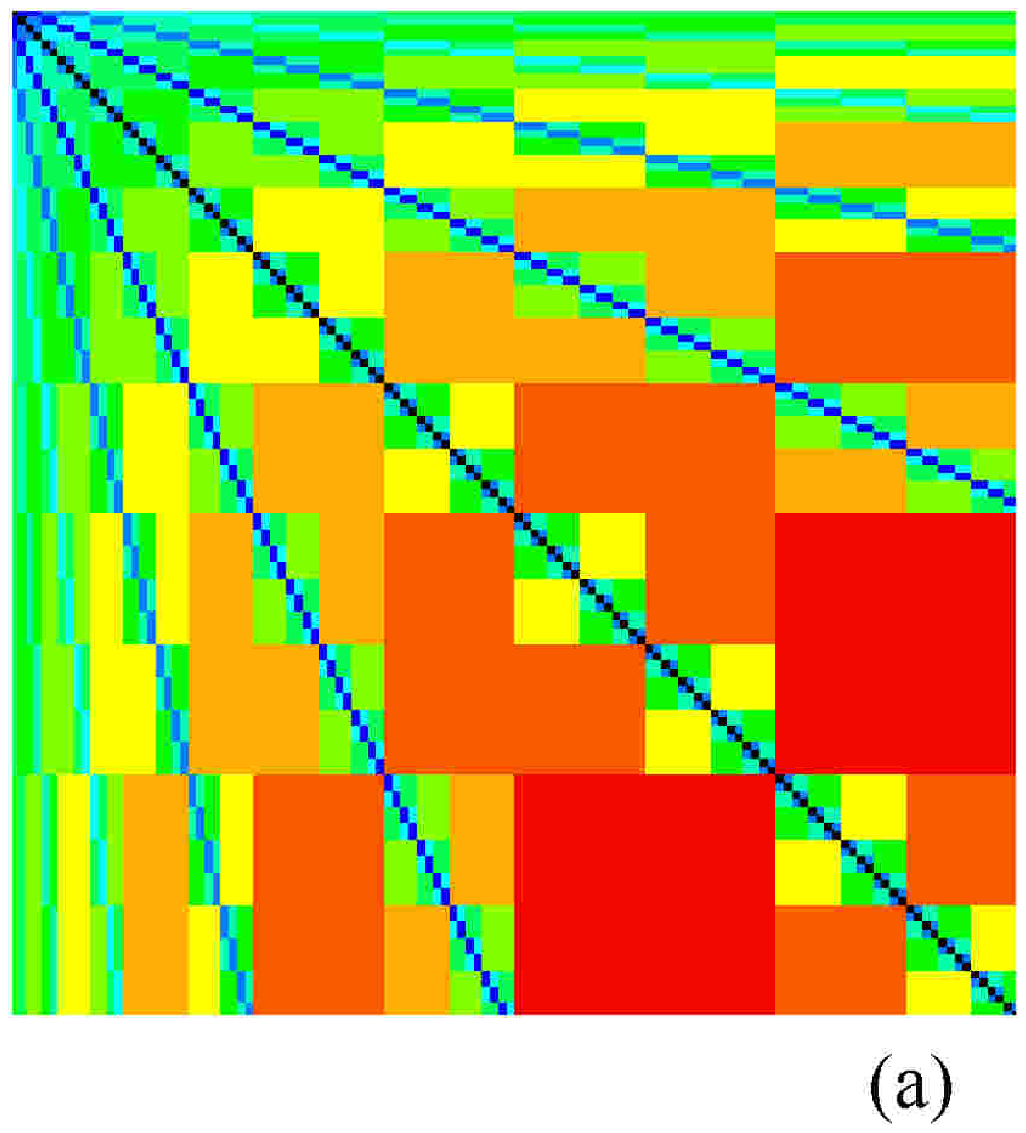}
\includegraphics*[width=8.cm,height=5.8cm,angle=0]{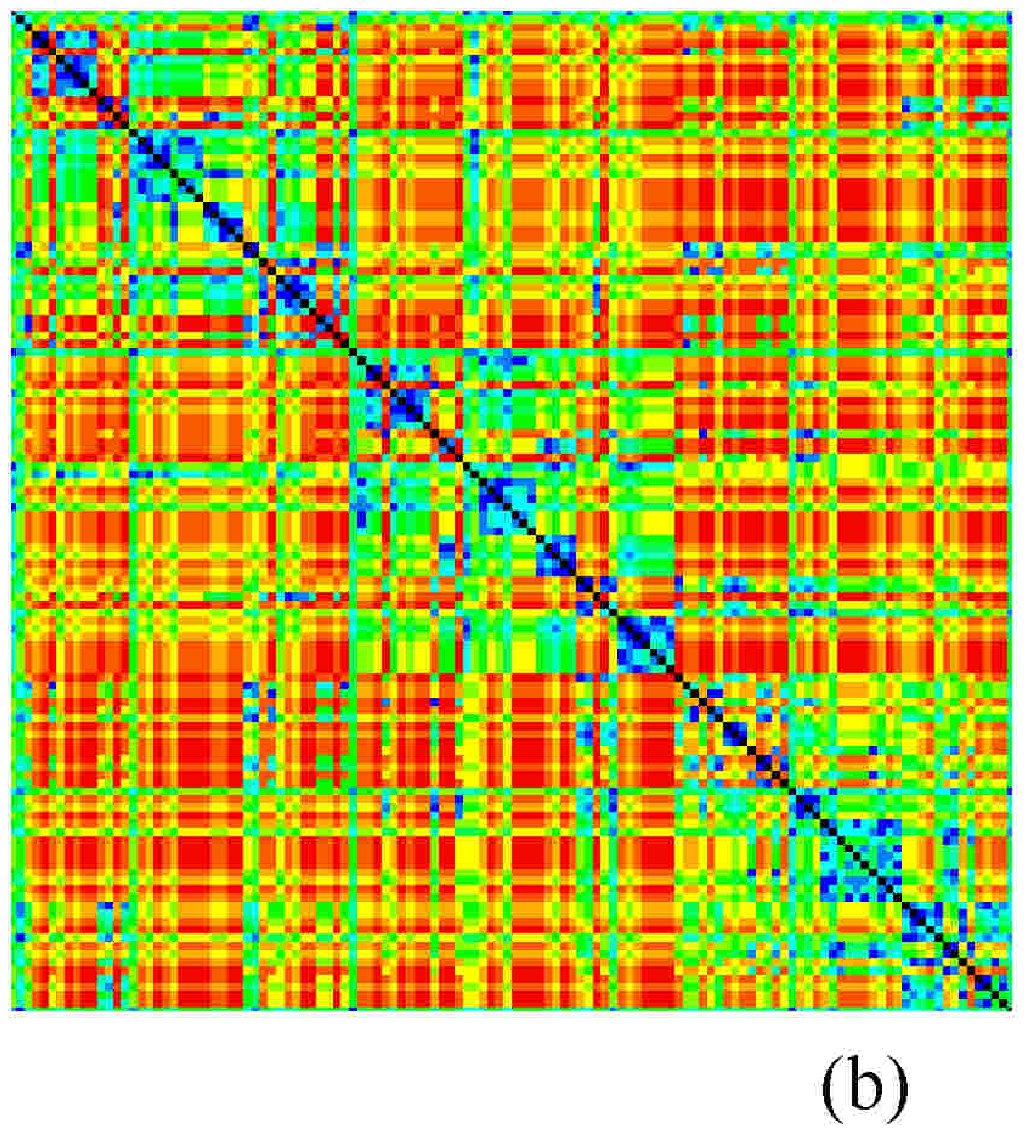}
\includegraphics*[width=8.cm,height=5.8cm,angle=0]{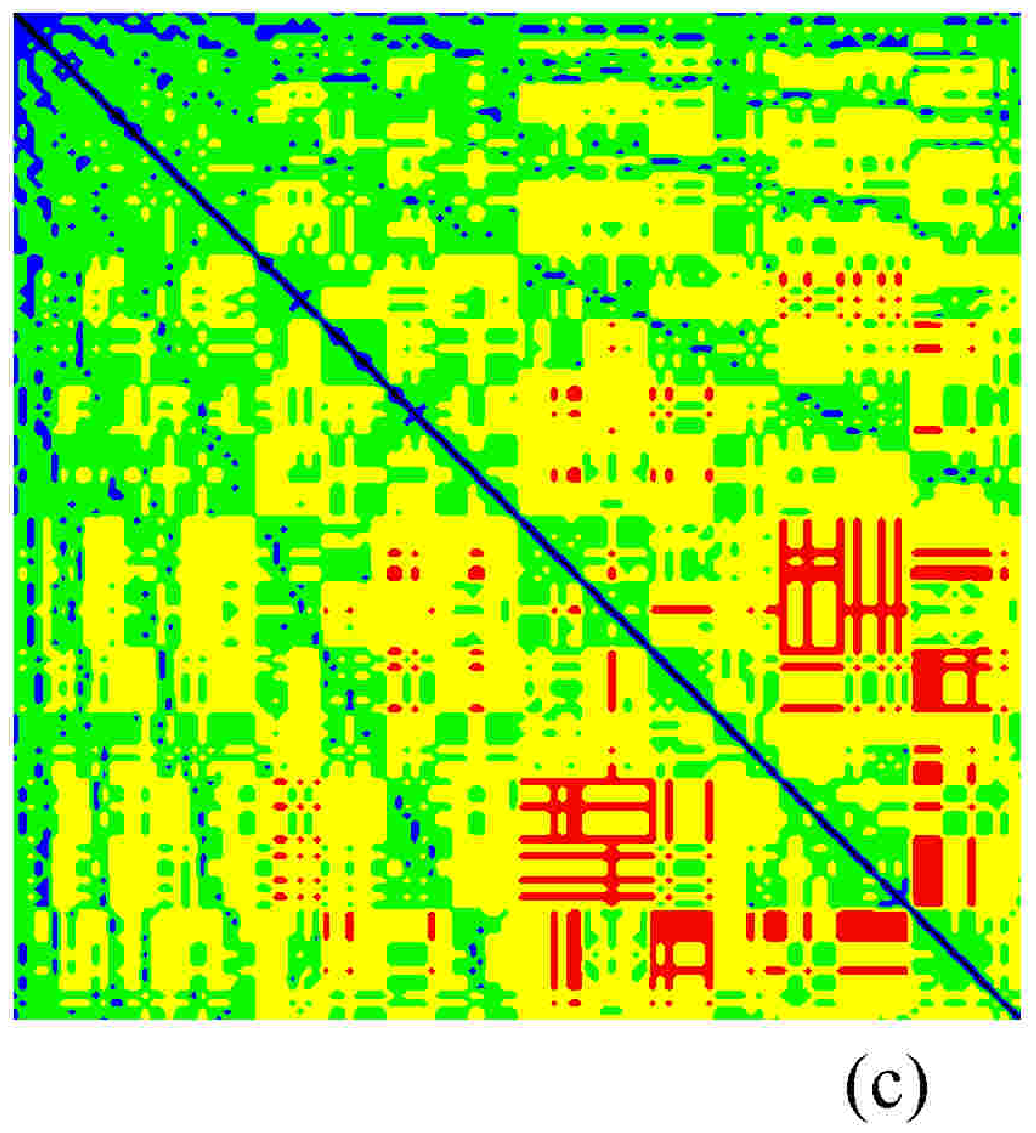}
\end{center}
\caption{Color representation of $\mathbf{\widehat{M}}$ for a number
of networks. As in Figure 8, $N=124$. In (a), (b) and (c), patterns
for, respectively $\mathbf{\widehat{M}}_{CT}$,
$\mathbf{\widehat{M}}_{CT \rightarrow AN}$, and
$\mathbf{\widehat{M}}_{AN \rightarrow CT}$. $D_{AN}=4$,
$D_{CT}=14$.} \label{fig9}
\end{figure}

The renumbering algorithms have then been applied to a large number
of pairs of matrices $\alpha$ and $\beta$, for which we also
generate the corresponding $\mathbf{\widehat{M}}_{\beta \rightarrow
\alpha}$ and $\mathbf{\widehat{M}}_{\alpha \rightarrow \beta}$. In
Figure 9(a-c) we illustrate how this scheme works in the cases of
the AN and the CT: In (a), (b) and (c) we draw, respectively, the
original $\mathbf{\widehat{M}}_{CT}$, in which the nodes are
numbered in concentric circles, starting from any arbitrary point,
$\mathbf{\widehat{M}}_{CT \rightarrow AN}$, where the same numbering
scheme in Figure 8a is used, and $\mathbf{\widehat{M}}_{AN
\rightarrow CT}$.

A first comparison of Figures 8(b) and 9(a) shows that the neighborhood
structures of AN and CT are indeed much more alike than one could infer by
visual comparison between Figures 8(a) and 9(a). In particular, diverging
rays from the upper left corner corresponding to low values of $\ell$ is a
common feature to both networks. Further comparisons among the two figures
indicate that the renumbering procedure causes a migration of the original
patterns to the one of the network it is projected on. Such fine details,
as changes in the angles formed by the diverging rays, are identified when
we also consider Figure 9c. It shows that the angles have changed with
respect to those in Figure 8b, following the values present in Figure 9a.
Similar evidences from the success of the renumbering CT according to the
original numbering in \cite{Andrade_spectrum} is provided by the
comparison of Figures 8a and 9b. All these effects are present despite the
fact that the diameters differ considerably ($D_{AN}=4$, $D_{CT}=14$), the
later requiring more colors/gray levels to represent each distinct value
of $\ell$.

These examples illustrate, in a very clear way, the discussion, in
the previous Section, on the contributions to distance between
networks resulting from adopted numbering and topological nature.
Finally, it should be remarked that the obtained value for
$\mathfrak{d}^{min}(AN,CT)$ does not depend on the initial numbering
of the minimization process, apart of typical fluctuations in the MC
and deterministic processes.

The values obtained for $\mathfrak{d}^{min}(\alpha,\beta)$ depends
on the different kinds to which $\alpha$ and $\beta$ belong.
Besides this obvious dependence, $\mathfrak{d}^{min}$ may also
depend on the number of nodes $N$, as well as on values assumed for
some parameter defining networks, as $p_a$ and $p_r$. Therefore, to
render the discussion clearer, we will discuss the effect of these
factors in separate.

To investigate the dependence of $\mathfrak{d}^{min}$ on network
sets, we first consider the completely ordered LC, together with the
AN, CT and SF networks. Four distinct samples, obtained from the
same random algorithm with distinct seeds, are chosen to represent
the SF network, while presented results are averages taken over the
distinct samples. Initially we analyze networks that are small in
size ($N=124$), but the resulting values for $\mathfrak{d}^{min}$
are typical for networks in these sets, as we shall see later on.
The results in Figure 10a represent the absolute minimum found over
a series of renumbering experiments. They reveal that
$\mathfrak{d}^{min}$ assumes smallest values for the (SF,SF) pair,
followed by (AN,SF), (AN,CT) and (SF,CT). As expected, LC lies at
much largest distance from the other three, specially with respect
to the CT. For this particular combination,
$\mathfrak{d}^{min}\simeq 0.45$ reaches the largest value in the
whole investigation. The entry (SF,SF) indicates the average value
over the 6 possible distinct pairs that can be formed from the 4
networks. We observe that, although they are built by the same
algorithm, the average distance between them is non-zero and
comparable to that for the (AN,CT) pair. We notice that, while the
already quoted networks quantifiers accurately express statistical
similarity among members of a same network set, they are not able to
uncover expressive differences among the actual neighborhood
structures for distinct samples within the SF network set, as the
analysis of $\mathfrak{d}$ does. On the other hand, a relative small
value for $\mathfrak{d}^{min}(AN,CT)$ goes along with several other
properties of AN, which are found to be quite similar to those of SF
networks.

The effect of $N$ in the values of $\mathfrak{d}^{min}$ can be checked in
the same set of networks, as none of them depend on any tuning parameter.
Therefore, in Figure 10b we show that the increase in the value of $N$ to
1096 only slightly affects the previous results. The measure
$\mathfrak{d}^{min}$ stays almost invariant with respect to the increase
in $N$, and also with respect to which sample has been used to represent
the SF network. This shows that $\mathfrak{d}^{min}$ is reliable in
capturing the geometrical and topological differences among the distinct
networks, so that it can indeed be used as a measure of their distance.

\begin{figure}
\begin{center}
\includegraphics*[width=6.6cm,height=4.73cm,angle=0]{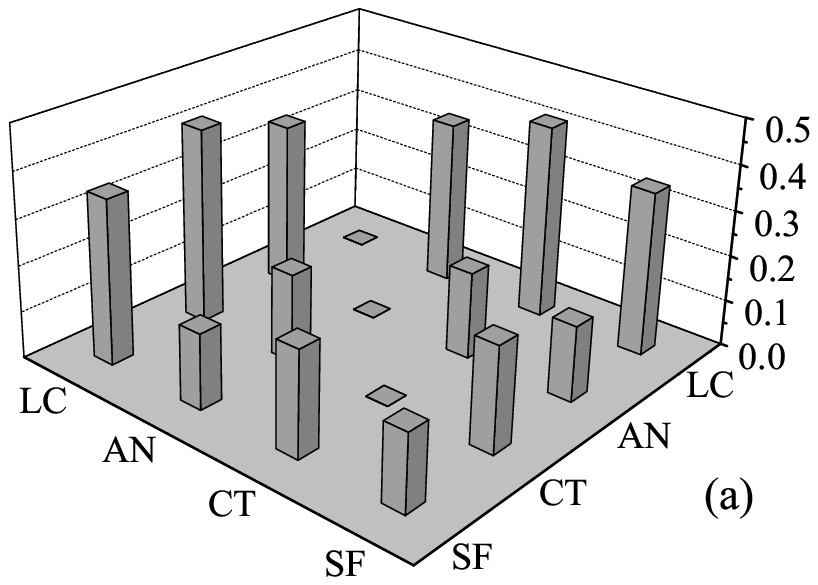}
\includegraphics*[width=6.6cm,height=4.73cm,angle=0]{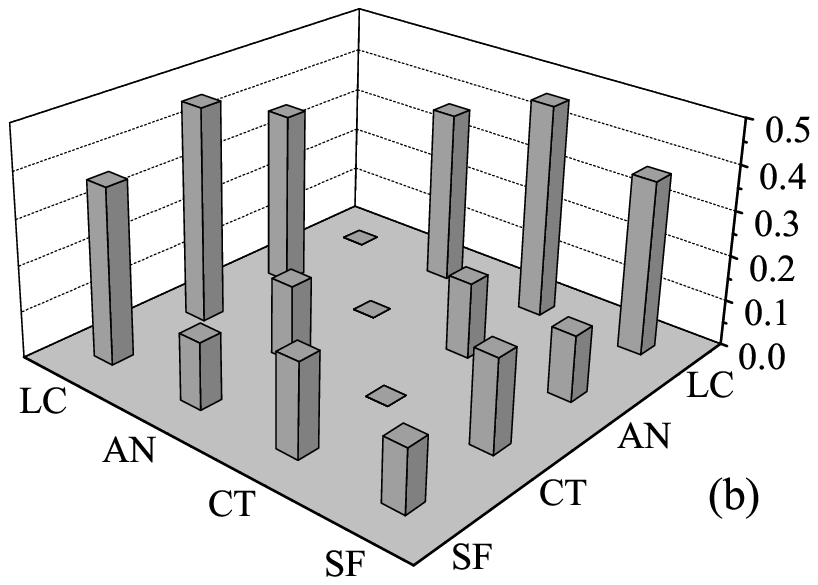}
\includegraphics*[width=6.6cm,height=4.73cm,angle=0]{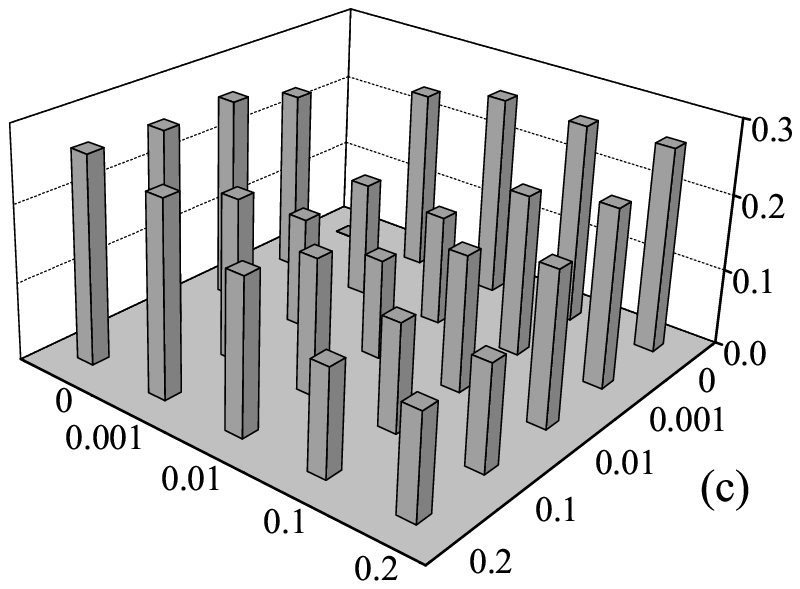}
\includegraphics*[width=6.6cm,height=4.73cm,angle=0]{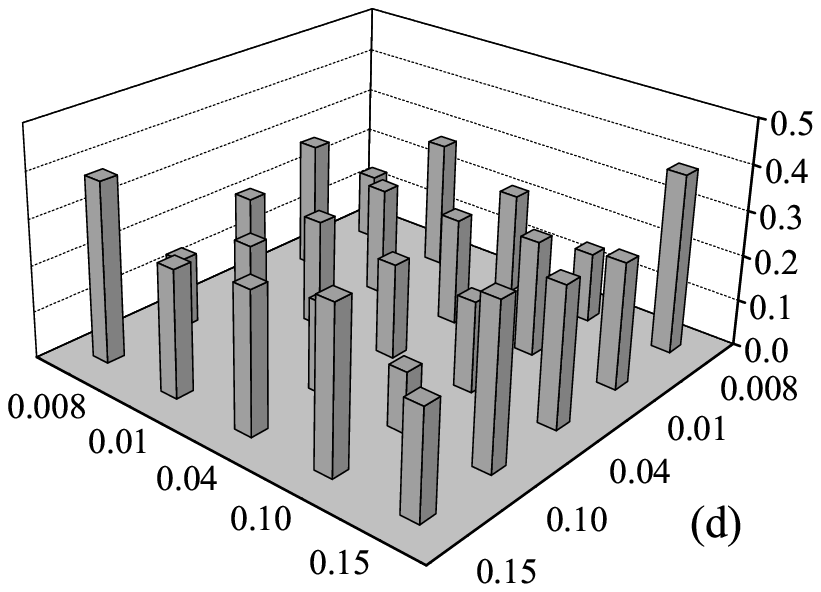}
\end{center}
\caption{Representation of $\mathfrak{d}^{min}$ by 3D bars (vertical axis)
for distinct situations. (a) $N=124$, and four distinct networks, as
indicated in the x and y axis labels. (b) The same networks, but for
$N=1096$. (c) and (d) Axis labels indicate five distinct values of $p_r$
and $p_a$, used in the analysis of the SW and ER network sets, with
$N=1000$.} \label{fig10}
\end{figure}

To investigate distances among networks in the same set, but with
distinct values of control parameter, we consider sets of SW and ER
networks. In Figures 10 (c) and (d) we show the behavior of
$\mathfrak{d}^{min}$ for distinct values of $p_r$ and $p_a$, holding
fix the value of $N$.

For the purpose of a better comparison, we also include the value
$p_r=0$ into the set of investigated SW networks. The results for
the largest value $p_r=0.2$ is typical for all non-zero values of
$p_r$. It assumes minimal values along the diagonal, i.e. when
$p_{r,\alpha}=p_{r,\beta}$ and increases when one moves away from
it, enlarging the difference between $p_{r,\alpha}$ and
$p_{r,\beta}$. The overall feature of (10c) is that of a valley
along the main diagonal. In a quantitative way, typical values in
the diagonal are given by $\mathfrak{d}^{min}\sim 0.14$, what is
somewhat smaller than that obtained for the SF
($\mathfrak{d}^{min}\sim 0.15$) shown in (10a) and (10b). The
maximal value $\mathfrak{d}^{min}$ in the investigated range of
$p_r$ is 0.27 for $\mathfrak{d}^{min}(p_r=0,p_r=0.2)$. Note that
this value is some 20\% larger than that found for
$\mathfrak{d}^{min}(SF,CT)$. Therefore, we see that pairs of
networks grown according to different rules can produce smaller
values of $\mathfrak{d}^{min}$ than pairs of networks originated by
the same procedure, whenever distinct values of one or more control
parameters, which one might guess to be more alike.

Finally, the results in Figure 10d for the ER set, where the range
of values for $p_a$ is similar to that used for $p_r$, goes in other
direction in comparison to SW. It is difficult to recognize a clear
pattern, reinforce the random character of this network class.
Values for $\mathfrak{d}^{min}$ along the diagonal are also among
the smallest, but near diagonal pairs can also show similar
distances. If we fix to the largest value of $\alpha$, and let
$\beta$ vary, no monotonic behavior is found as before. From the
quantitative point of view, the mean value along the diagonal
$\simeq 0.19$ is higher than those for SW or SF. The maximal
off-diagonal value $\simeq 0.39$ is more than 30\% higher that the
corresponding value for the SW set.

\section{Conclusions}

In this work we put forward, in a systematic way, the investigation
of the higher order neighborhoods of a complex network $R$. The
basic concepts, that have been introduced in a previous work, are
developed in many aspects, opening new paths for the exploration of
properties related to structures in a larger scale than that of the
immediate vicinity of a node. According to the basic idea, each
neighborhood $O(\ell)$, represented by an adjacency matrix
$M(\ell)$, is regarded as a network in itself, so that many distinct
quantifiers used for network characterization can be used to obtain
the properties of each neighborhood.

The large amount of results we obtain can be casted into two parts. The
first one encompasses all results obtained from each $M(\ell)$, while the
second is related to properties of $\mathbf{\widehat{M}}$, which
condenses, in a single matrix, all information on the neighborhood
structure.

In the discussion of the distinct quantifiers, we find that many of
them behave in a oscillatory way with respect to $\ell$. This is the
case of $C(\ell)$ for self-similar networks, as CT, DHL, WHL, as
well as for hypercubic lattices. For ER and SW networks, oscillatory
behavior in $C(\ell)$ can also be found, provided linking and
rewiring probabilities are small enough. Interesting enough, the
values of $\langle k(\ell)\rangle$ oscillate for both DHL and WHL,
for all $\ell$ range. For all other networks, the average node
degree $\langle k(\ell)\rangle$ increases exponentially for several
networks, but this behavior can be masked, for networks with very
small values of $D$. Hierarchical property and scale-free
distribution of nodes are rare events for large values of $\ell$.
Nevertheless, we present examples of SF, AN and WHL, where
corresponding power law distributions for $p(k;\ell)$ and
$C(k;\ell)$ are found for $\ell>1$. Properties related to degree
assortativity or dissortativity are likely to remain the same as
$\ell$ increases. One exception, again due to the emergence of
oscillatory behavior, refers to the self similar DHL and WHL.
Finally, we use the information obtained from the set ${M(\ell)}$ to
evaluate the fractal dimension of all investigated networks. The
obtained results reproduce well into the expected values, among
which we quote $d_F\rightarrow\infty$ for CT, smaller values of
$d_F$ for less connected networks, and some troubles in finding
precise values when the network diameter is very small.

The properties of $\mathbf{\widehat{M}}$ can be used for a variety
of purposes. The simple visualization of $\mathbf{\widehat{M}}$ with
the help of color codes can be very helpful for identifying and
understanding the large scale structure of networks. Such procedure
naturally evolves to the question of finding which representation,
depending on the node numbering adopted, is more suitable to be used
for a given network. Examples for the deterministic AN illustrate
how distinct they can be. Furthermore, since a minimization
Monte-Carlo algorithm finds the best way a given network can be
\textit{projected} onto a second one, we address the issue of
defining and evaluating the distance between networks based on the
neighborhood structure contained in $\mathbf{\widehat{M}}$. We
present results that support, in quantitative and qualitative way,
the usefulness of this framework. We can assign quantitative values
for a distance function, which corresponds to several intuitive
notion of distinct structures, as that between LC and CT. The
results we obtain also reveal that, despite the fact that the
members of a set of networks, generated by the same algorithm, share
the same values of the usual network indices, their actual
neighborhood structure can remain very distant from each other,
specially for the ER situation.

This series of investigations we open in this work is very large and can
be explored in many directions. They certainly include a finer exploration
of sets of well known networks as well as analyzes of data from actual
networks. Efforts to extract further information along new lines from the
matrices $M(\ell)$ and $\mathbf{\widehat{M}}$ are on consideration as
well.

\textbf{Acknowledgement:} This work was partially supported by
CNPq and FAPESB.

\bibliographystyle{prsty}

\end{document}